\documentclass[letterpaper,twocolumn,10pt]{article}
\usepackage{usenix2019_v3}

\usepackage{tikz}
\usepackage{amsmath}
\usepackage{authblk}
\usepackage{filecontents}
\usepackage{xspace}
\usepackage{enumitem}

\usepackage{color}
\usepackage{subcaption}
\usepackage{algorithm}
\usepackage{algorithmic}
\usepackage[skip=4pt]{caption}
\usepackage[medium, compact]{titlesec}

\renewcommand{\algorithmiccomment}[1]{\bgroup\hfill \textbf{\color{red}$\rightarrow$ ~#1}\egroup}

\usepackage{lipsum}

\def\commentsDisplay{}
\def\ArxivVersion{}

\ifdefined\commentsDisplay
    \newcommand{\aditya}[1]{\textbf{\color{blue}AA: #1}}
    \newcommand{\minlan}[1]{\textbf{\color{red}Minlan: #1}}
    \newcommand{\lam}[1]{\textbf{\color{brown}Lam: #1}}
    \newcommand{\jiaqi}[1]{\textbf{\color{green}Jiaqi: #1}}
    \newcommand{\todo}[1]{\textbf{\color{cyan}TODO: #1}}
    \definecolor{airforceblue}{rgb}{0.36, 0.54, 0.66}
\else
    \newcommand{\aditya}[1]{}
    \newcommand{\minlan}[1]{}
    \newcommand{\lam}[1]{}
    \newcommand{\jiaqi}[1]{}
    \newcommand{\todo}[1]{}
\fi

\ifdefined\highlightvisvion
    \newcommand{\revision}[1]{{\color{blue}{#1}}}
\else
    \newcommand{\revision}[1]{{#1}}
\fi

\newcommand{\secref}[1]{\S\ref{#1}}

\usepackage{tikz}

\newcommand{\mypar}[1]{\vspace{0.00cm} \noindent \textbf{#1}}

\begin{document}
\date{}

\ifdefined\ArxivVersion
    \title{EdgeSight: Enabling Modeless and Cost-Efficient Inference at the Edge}
    \newcommand{\sysname}{EdgeSight\xspace}
\else

    \title{DragonEye: Cost-efficient Modeless Serving at the Edge}
    \newcommand{\sysname}{DragonEye\xspace}
\fi

\author{
  ChonLam Lao$^{\ddagger}$, 
  Jiaqi Gao$^{\ddagger}$, 
  Ganesh Ananthanarayanan$^{\S}$, 
  Aditya Akella$^{\diamond}$, 
  Minlan Yu$^{\ddagger}$ 
}

\affil{\textit{$^{\ddagger}$Harvard University, 
  $^{\S}$Microsoft, 
  $^{\diamond}$University of Texas at Austin}} 


\maketitle
\begin{abstract}

Traditional ML inference is evolving toward \textit{modeless inference}, which abstracts the complexity of model selection from users, allowing the system to automatically choose the most appropriate model for each request based on accuracy and resource requirements.  While prior studies have focused on \textit{modeless} inference within data centers, this paper tackles the pressing need for cost-efficient \textit{modeless} inference at the edge---particularly within its unique constraints of limited device memory, volatile network conditions, and restricted power consumption.

To overcome these challenges, we propose \sysname, a system that provides cost-efficient \textit{modeless} serving for diverse DNNs at the edge. \sysname employs an edge-data center (edge-DC) architecture, utilizing confidence scaling to reduce the number of model options while meeting diverse accuracy requirements. Additionally, it supports lossy inference in volatile network environments. Our experimental results show that \sysname outperforms existing systems by up to $1.6\times$ in P99 latency for modeless services. Furthermore, our FPGA prototype demonstrates similar performance at certain accuracy levels, with a power consumption reduction of up to $3.34\times$.

\end{abstract}
\pagestyle{plain}
\section{Introduction}

Edge-based machine learning inference has become increasingly important in powering modern applications like Automatic License Plate Recognition (ALPR)~\cite{alpr1,alpr2}, road object detection and classification~\cite{recl, ekya}, surveillance intruder detection~\cite{dds_sigcomm}, and other real-time systems~\cite{gemel, adainf}. To reduce latency, these services are often deployed on edge servers strategically positioned closer to end-users, typically between cellular networks and data centers (Figure~\ref{fig:edge_dc_deployment}). This proximity minimizes computational overheads, such as preprocessing costs for inference, enabling faster and more efficient service delivery.


As edge services continue to evolve, the application and user demands have grown increasingly diverse, even when leveraging the same underlying model with varying accuracy requirements~\cite{INFaaS, mlperf}. For instance, object detection in smart factories requires \textit{high accuracy} for robotic arms to handle small components, as errors could damage equipment or produce defects. In contrast, edge security systems monitoring low-risk areas can tolerate \textit{lower accuracy}, as minor detection errors can be mitigated through additional verification. Additionally, differentiated pricing models and user tiers (e.g., free versus enterprise users) have created the need to support varied accuracy requirements.

These demands are driving a shift from traditional inference to \textit{modeless} prediction serving~\cite{INFaaS, clipper, sommelier}. In this paradigm, developers pre-register models with varying accuracy requirements into the serving system.   hen users submit inference requests, the system dynamically selects the most appropriate model to meet the specific requirements~\cite{mlperf}. The trend toward modeless serving offers finer granularity for users and service providers, reducing service costs while meeting users' specific needs. A serving system must address these diverse requirements while minimizing costs.

Yet, building such \textit{modeless} inference systems for edge serving support is not easy. Unlike traditional data centers~\cite{kim2023greenscale}, edge prediction serving systems must operate in resource-constrained environments with strict limitations on power and budget~\cite{gemel, ekya, dds_sigcomm, microsoft-blog-cascading-inference, microsoft-blog}. While prior works~\cite{INFaaS, clipper, modelswitch} have explored modeless serving in cloud settings and demonstrated benefits such as lower latency and reduced deployment costs, edge deployments for modeless serving remain scarce. This scarcity stems from several key challenges:

\begin{figure}[t]
    \centering
    \includegraphics[clip, trim=0cm 11cm 7cm 0cm, width=1\linewidth]{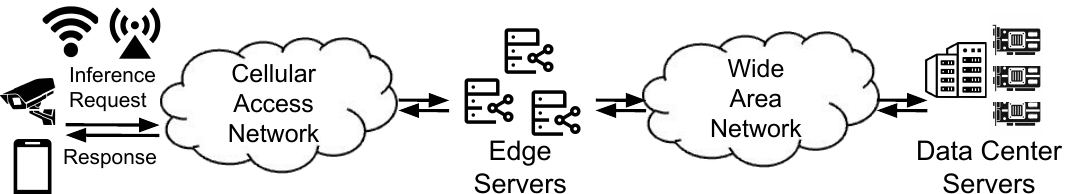}
	\caption{Edge-DC Deployment}
    \label{fig:edge_dc_deployment}
    \vspace{-6mm}
\end{figure}


 
\textbf{Challenge \textcircled{1}: Serving a wide range of accuracy requirements at the edge is expensive}. 
To meet diverse accuracy requirements from users, serving system needs to handle a sufficient number of model options from developers. Otherwise, picking an overkill model for an application with low accuracy requirement incurs unnecessary inference computational costs. MLPerf~\cite{mlperf} shows that choosing a model with slightly higher accuracy (e.g., a few percent) can lead to 5-10x computational costs. Therefore, it is important to host a wide range of models for various accuracy needs and dynamic workloads. 

However, hosting many models concurrently is impossible given the limited memory at edge GPU and CPU~\cite{gemel, ekya}. Therefore, we have to load pre-registered target serving model options in the runtime~\cite{clockwork, gemel, modelswitch, mark}, but frequent runtime model loading (cold-start inference) between GPU and CPU can introduce unpredictable latency and result in low GPU utilization. Recent research indicates that the load time versus the inference time can be up to 3:1~\cite{clockwork}.

\textbf{Challenge \textcircled{2}: Tolerating high network volatility in edge serving.}
Edge network is often unreliable due to poor network quality caused by multi-tenancy and bursty behavior in the wild, leading to unpredictable and high tail latency due to queuing or retransmission at the edge~\cite{zhou2024augur, hairpin}. Moreover, real-world inference-serving workloads exhibit burstiness with extremely high short-term variance~\cite{SHEPHERD, mark, modelswitch}, putting more pressure on the edge network during serving. These network challenges significantly contribute to high end-to-end inference times (as discussed in Section \secref{mov:network_tax}).



\textbf{Challenge \textcircled{3}: Serving ever-growing processing costs with limited power budget.} Unlike data centers, edge inference systems have significantly fewer resources and power supplies~\cite{kim2023greenscale, recl, gemel}. At the same time, even a low-end GPU (e.g., Nvidia T4) can consume up to 75W, which creates pressure for widespread deployment in edge environments. Inference processing tasks are becoming increasingly complex, encompassing tasks such as JPEG decoding and other image preprocessing functions, which contribute to nearly 32-59\% of the end-to-end inference latency (as discussed in Section \secref{mov:data_processing}). Achieving both \textit{low latency} and \textit{low power consumption} simultaneously can be challenging.

To address these challenges, we propose \sysname, which provides low-latency modeless serving of vision DNN. 
There are three key designs in \sysname:

First, we propose \textit{confidence scaling}. Instead of switching models at runtime for different target models~\textcircled{1}, we employ a dual-model approach to navigate the entire accuracy space. When a request is received, it is sent to two models concurrently: a lightweight model serves as the frontend to provide a fast answer and obtain the output 'confidence score.' If the confidence is high (deemed trustworthy), the response is returned directly to the user, and the backend is notified to stop the inference.

Second, to reduce network delay in an unreliable network environment~\textcircled{2}, we propose a hardware-friendly \textit{unreliable} transport layer at the edge that allows \textit{lossy compressed format (e.g., JPEG) recovery} in a streaming manner when packet loss occurs. This enables inferences with partial information (e.g., inference on images with missing pixels) to reduce latency and avoid retransmission costs. Combined with \textit{confidence scaling} and fallback mechanisms, we do not need to sacrifice accuracy.

 

Third, we built two prototypes, one GPU-based deployment-friendly frontend prototype and a cost-effective FPGA frontend prototype. The FPGA prototype not only hosts the lightweight model but also supports the entire inference pipeline in a streaming-friendly manner, including packet processing, the network stack, and data preprocessing as a holistic solution, leveraging the architectural benefits of FPGA. Our FPGA prototype further verifies the feasibility of hardware implementation of both \textit{confidence scaling} and \textit{lossy compressed format recovery}, demonstrating a low-latency and high-power-efficiency solution compared to GPU-based accelerators.
Our evaluation shows that \sysname achieves up to $1.37\times$ improvement in P90 latency and $1.6\times$ improvement in P99 latency compared to Clockwork serving modeless in a multi-tenancy environment with different hardware profiles. Our FPGA also demonstrates up to $1.5\times$ and $3.34\times$ reductions in power consumption while serving the same 75\% accuracy requirement.

\vspace{-1mm}
\section{Motivation and Challenges}


Edge modeless serving (\secref{mov:model_serving}) aims to offer applications with various user-specified accuracy requirements (\secref{mov:runtime_cost}) in environments with poor network conditions (\secref{mov:network_tax}) and the limited resources and power constraints (\secref{mov:data_processing}), posing challenges to the underlying serving system.


\revision{

\subsection{Modeless Serving}
\label{mov:model_serving}
As inference serving matures, the trend of hiding model details and allowing users to express high-level intentions for inference is becoming increasingly popular~\cite{clipper, Cocktail, sommelier, INFaaS, ahmad2024proteus}. This is a strong requirement for both developers and service providers (e.g., AWS). Developers aim to offer maximum flexibility, enabling various users and applications to utilize their models in different ways~\cite{INFaaS, mlperf}, or by building service tiers to support different business models. Meanwhile, service providers seek more fine-grained control over model selection, enabling better resource scheduling and maintaining quality of service, such as handling request bursts~\cite{SHEPHERD}, to reduce serving costs.

However, these serving systems, which provide inference with differing requirements, such as accuracy, face a common challenge: the system needs to assemble different models to create an accuracy spectrum~\cite{clipper, Cocktail, sommelier, INFaaS, ahmad2024proteus}, especially when service providers are not permitted to access or modify developers' models due to privacy concerns. Unfortunately, this approach results in high inference service costs.
}

\subsection{Serving Applications with Diverse Accuracy Requirements}
\label{mov:runtime_cost}

\begin{figure*}[t]
    \centering
    \begin{minipage}{0.9\linewidth}
        \begin{minipage}{0.33\linewidth}
            \centering
            \includegraphics[width=\linewidth]{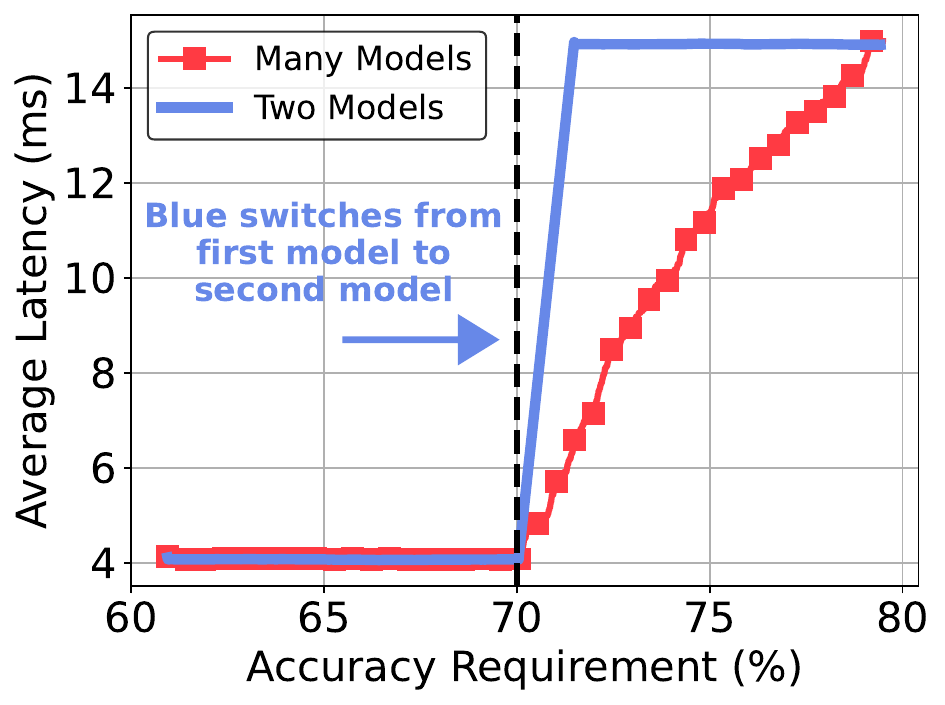}
            \caption{Serving different requirements with two models}
            \label{fig:granularity}
        \end{minipage}\hfill
        \begin{minipage}{0.33\linewidth}
            \centering
            \includegraphics[width=\linewidth]{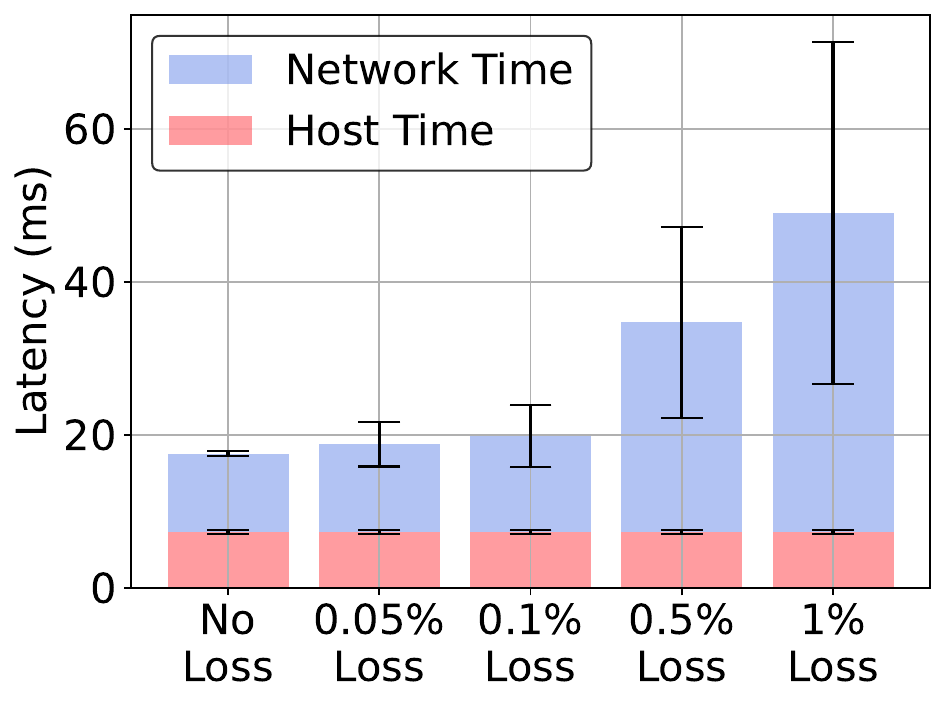}
            \caption{Impact of Packet loss}
            \label{fig:internet_loss_time}
        \end{minipage}\hfill
        \begin{minipage}{0.33\linewidth}
            \centering
            \includegraphics[width=\linewidth]{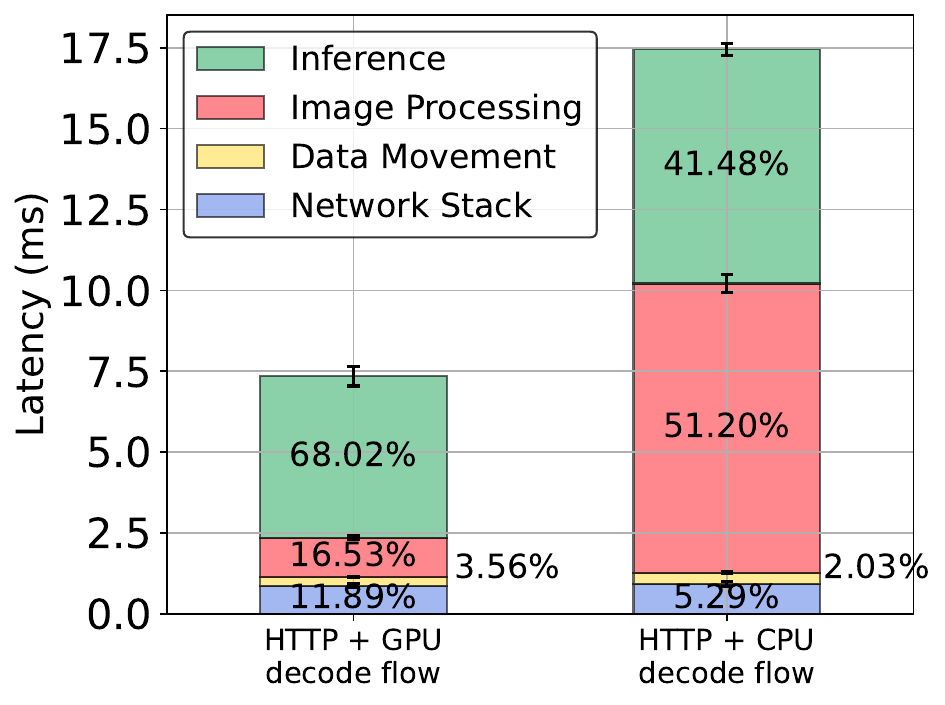}
            \caption{Breakdown of inference latency with 0ms network delay}
            \label{fig:breakdown}
        \end{minipage}
    \end{minipage}
    \vspace{-6mm}
\end{figure*}




Selecting the right model variant with the appropriate accuracy is crucial for modeless serving cost~\cite{mlperf, INFaaS}. However, this presents a two-sided challenge. 

On one end, serving a massive amount of model options with varying requirements introduces high swapping costs during runtime. Even in image classification tasks alone, there are at least 44 vision models available, each offering varying trade-offs between computation and accuracy within an accuracy range of 55\% to 83\%\cite{mlperf}. Since GPU memory is expensive, models cannot be stored in GPUs indefinitely and therefore require frequent swapping.
However, frequent loading and unloading of models during runtime can result in unpredictable latency. Clockwork\cite{clockwork} shows that the model transfer time between GPU and CPU can be up to 3X the model inference time. This runtime cost accumulates in the client end-to-end latency if the model is not loaded (\textit{cold inference}). Moreover, real-world workloads are highly dynamic~\cite{SHEPHERD, alpaserve}, and it is difficult to predict the workloads accurately. Incorrect estimation of the demands for specific models can lead to significant runtime GPU resource scaling and allocation issues.

On the other end, failing to provide a sufficient number of models to express the spectrum causes a lack of granularity in model selection, which results in using overkilled models to address simpler demands, leading to unnecessary inference costs. 

We demonstrate the drawback of using an overkill model to serve requests with different accuracy levels in an experiment (Figure~\ref{fig:granularity}). We modify Clockwork~\cite{clockwork}, a state-of-the-art serving system, to support modeless. We pre-register two models provided in the Clockwork repository into the serving system to accommodate different accuracy requirements, namely $winograd\_resnet18\_v2$ (with 70.65\% accuracy) and $resnest101$ (with 82.17\% accuracy).  We use this setup to compare with another run where a massive amount of models with different accuracy levels are loaded into the system to serve varying accuracy requirements. This experiment does not incur any swapping latency.


We initialize 4000 requests with varying accuracy requirements, ranging from 60\% to 80\%, in ascending order, and send them to the serving system. The serving system selects the model that can provide sufficient accuracy for each request. In the original system, with around a 70\% accuracy requirement, the average latency significantly increases due to the serving system switching to an overkill model. The result shows that the use of many models provides better model allocation, resulting in improved average latency than using two models.

\subsection{Fluctuating Network Conditions on Edge}
\label{mov:network_tax}




High tail latency and packet loss are common issues encountered in edge networks due to the increasing trend of multi-tenancy and bursty behavior during peak usage times~\cite{zhou2024augur, hairpin, imc17}.


Tail latency is the primary cause of performance reduction. AUGUR~\cite{zhou2024augur} demonstrates that in edge settings, long latency is observed, with the median round-trip time (RTT) staying below 30ms, while the 99.9th percentile latency exceeds 200ms. These long tail latencies stem from RTT inflation induced by fluctuations in edge wireless paths. Another network problem is edge packet loss~\cite{hairpin}. This occurs when the number of users is high or the quality of the wireless environment is poor~\cite{imc17}, and it typically dominates the end-to-end delay of inference.

To demonstrate how packet loss affects inference performance in edge, we conducted an inference task on the ImageNet dataset~\cite{imagenet} using the state-of-the-art ConvNext model~\cite{convnext} on the Nvidia A100 GPU. Figure~\ref{fig:internet_loss_time} illustrates the impact of network loss on inference time with a 10ms network delay and loss rates ranging from 0.05\% to 1\%.
With 1\% packet loss, the inference time increases by $4\times$ compared to no packet loss. This delay occurs because it takes several round-trips to recover lost packets before the full image can be decoded and inference can begin. Notably, the network time significantly exceeds the computation time.

\subsection{High Performance Inference Stack}
\label{mov:data_processing}

Apart from inference, other processing tasks contribute a significant portion of the end-to-end delay. 

In Figure~\ref{fig:breakdown}, we measure the breakdown of the average inference latency at the host in an image classification task with the same setting as in Section~\secref{mov:network_tax}, with network delay set to 0.
In the HTTP + CPU decoding workflow (the left bar), the inference time constitutes only about 41.48\%, with the remaining time consisting of \textbf{(1) HTTP/gRPC and TCP stack processing (5.29\%)}, which would further increase if we experience long network delays and/or packet loss, as demonstrated in~\secref{mov:network_tax}. \textbf{(2) Data movement (2.03\%)}, although it represents a small portion in the classification task, it is worth noting that it incurs a non-negligible cost when the inference task is communication-intensive or simple, as shown in \cite{n3ic}, due to multiple data transfers among NIC, CPU, and GPU. \textbf{(3) JPEG image decoding and preprocessing (e.g., cropping, centralizing) (51.20\%)}, as clients typically send data in a compressed format to reduce communication volume. 

If we optimize the JPEG decoding task using GPU accelerator~\cite{dali} (the right bar), the additional network stack and data processing time still account for approximately 32\% of the total inference time. It becomes evident that (1) the majority of time is spent on image processing and inference, making accelerators like GPUs play a crucial role, and (2) as inference and data processing are optimized, the network stack and data movement present more room for further improvement. However, in edge settings, high-performance GPU options are limited due to power supply constraints, and optimizing data movement and network transport is challenging due to these accelerator constraints (e.g., network transport is difficult to offload to the GPU, and the workflows between the NIC, CPU, and GPU are hard to streamline). Therefore, we need an acceleration solution that can deliver high processing performance while staying within power limits.





\vspace{-1mm}
\section{Overview}

To address the challenges of model deployment (\secref{mov:runtime_cost}), network (\secref{mov:network_tax}), and performance efficiency (\secref{mov:data_processing}), we propose \sysname. \sysname's goal is to provide a \textit{low end-to-end inference latency} serving system for applications with diverse accuracy requirements and user workloads. It also aims to \textit{minimize deployment resources} to best fit edge scenarios.

\begin{figure}[tb]
    \centering
    \includegraphics[clip, trim=0cm 2.8cm 0cm -2cm, width=\linewidth]{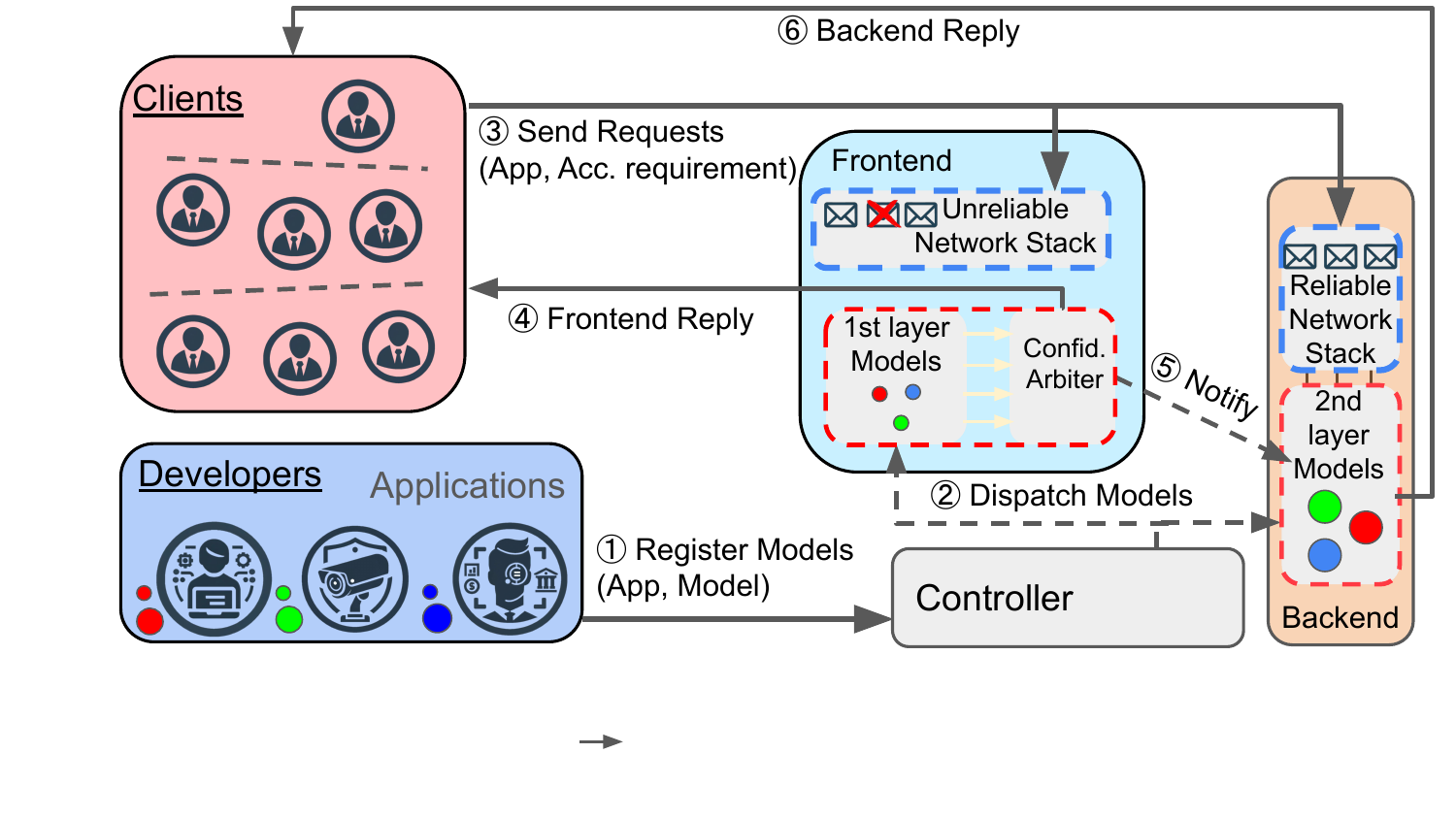}
	\caption{\sysname Overview}
    \label{fig:nic_overview}
    \vspace{-6mm}
\end{figure}



Our key insights of \sysname designs are as follow:

{\bf Hierarchical inference with confidence scaling:}
Applications have different accuracy requirements according to their user needs or pricing schemes. To serve a wide range of accuracy requirements at low cost, we employ a hierarchical architecture that utilizes two models for each application: the frontend includes small models, providing ultra-low latency but moderate-level accuracy to serve requests initially, and the backend uses large models with the best accuracy.
Both models process the request concurrently, if the frontend has high confidence in its response, it sends
a cancel to the backend model to reduce cost. 
Otherwise, the frontend drops its response and waits for the backend to propose a high accuracy result.
This approach allows us to reduce both the latency and cost needed to process requests with diverse application requirements without losing accuracy.





 
{\bf Unreliable transport that tolerates packet loss and high tail latency:} To reduce the network overhead associated with handling packet loss, reordering, or tailed packets, we propose running unreliable transport within the frontend. Our lossy inference design supports a hardware-friendly, streaming-based corrupted image recovery scheme to enable making inferences with partial information, tolerating packet loss, and disregarding tailed packets in a low-quality or long-distance Internet network connection.


The key benefit of such a frontend-backend design approach is that it has minimal impact on the original inference network protocols and inference models at backend hosts while allowing for innovation on the frontend design. This ensures compatibility with different inference frameworks. If the frontend fails, the original inference backend can still continue to work.

{\bf Holistic Streaming-Friendly FPGA Prototype:} 
We also demonstrate the feasibility of implementing the two core ideas above in a cost-effective, full-stack FPGA prototype. Data processing and inference overheads contribute significantly to the end-to-end time, and there is potential for improvement in both power consumption and performance with GPUs. By leveraging the specialization of FPGA architecture—such as efficiently handling streaming-friendly tasks like JPEG decoding and lossy network protocols, and avoiding multiple PCIe transmissions between the CPU, GPU, and NIC—latency can be reduced. Moreover, using FPGAs can be more cost-effective and power-efficient than GPUs, making them ideal for edge deployments where power and memory constraints are typically tight~\cite{blott2018finn, finn, DBLP:conf/icfpt/PetricaAKFCB20,DBLP:journals/corr/abs-2007-10451,10.1145/3404397.3404473,10.1145/3289602.3293915,8825027,10.1145/3330345.3330385,10.1145/3330345.3330385,10.1145/3289185,brainwave1, brainwave2,aws_f1_instance}.

Figure~\ref{fig:nic_overview} shows \sysname's workflow. There are two components in \sysname: the frontend and the backend. In the preparation phase, developers first need to register applications with two models in our controller \textcircled{1}: a lightweight model in the frontend and a powerful model in the backend. Once the dual models have been registered, the controller calculates the corresponding accuracy requirement with the required confidence thresholds, determining how confident the answers need to be to be directly sent back to the client by the frontend. When a user sends a request with a accuracy requirement, we map it to that confidence threshold for the request to achieve the desired end-to-end accuracy \textcircled{2}.


After the preparation phase, clients send inference requests for their applications with different accuracy requirements to both \sysname's frontend and backend \textcircled{3}. The frontend uses an unreliable network stack and a small model to handle packet loss, recover images, and perform lossy first-layer inference, while the backend uses a reliable network stack and a powerful model as the original setting. When the inference result is too difficult for the frontend (which might be due to a difficult request or the frontend's small model lacking confidence due to lossy inference), it falls back to the backend for inference.

If the \sysname frontend is able to handle the request with confidence, it directly sends the result back to the client \textcircled{4} and notifies the backend to terminate the ongoing inference \textcircled{5}. Otherwise, it waits for the backend inference to complete and sends the final result back to the client \textcircled{6}.

In the following sections, we discuss the use of confidence scaling (\secref{sec:confidence_scaling}) and our unreliable transport design (\secref{sec:unreliable_transport}) to achieve low deployment cost and low latency. We then introduce how we implemented these two ideas into our system and the FPGA prototype (\secref{sec:implementation}) to achieve high power efficiency.

\section{Confidence Scaling}
\label{sec:opportunites}
\label{sec:confidence_scaling}



In this section, we introduce how \sysname leverages the confidence score~(\secref{sec:confidence_score}) to cover the accuracy spectrum for modeless serving with the minimal cost, and explain how this method performs inference with request-level accuracy, followed by the justifications behind this approach~(\secref{sec:confidence_threshold}).

\subsection{Confidence Score as Threshold}
\label{sec:confidence_score}

\sysname introduces confidence scaling to resolve the two-end puzzle between the high model swapping rate and unnecessary inference latency overhead mentioned in Section~\secref{mov:runtime_cost}. The idea is to use the confidence score output from the lightweight model to estimate the confidence of the request's answer and determine if it should be sent to the backend model. This approach provides two benefits: (1) We can control the proportion of requests sent to the backend, creating a mechanism for achieving the entire spectrum of end-to-end accuracy requirements. (2) The correlation between high confidence and correct answers allows for more efficient labor division between the frontend and backend. This enables the frontend to handle requests directly when deemed trustworthy, resulting in an improved performance-cost curve.

There are different ways to determine the confidence of an answer given by the model. In \sysname, we use the last softmax layer of the CNN model and select the N-th highest probability as the result's confidence score when the inference job selects the top-N outputs. This confidence score is model-local, easy to obtain, and simple to calculate. Other methods of calculating the confidence score and different applications, such as language models, are discussed in \secref{dis:language}.


Previous work~\cite{microsoft-blog-cascading-inference} employ confidence scaling to handle requests in a cascading manner: a lightweight model handles easy requests, while difficult requests from the frontend are forwarded to a more powerful backend model when the confidence score is insufficient. Although the basic idea is similar, our use of confidence scaling aims to create an accuracy spectrum to minimize the serving system cost in response to the increasing demands for serving different accuracy requirements, rather than focusing solely on inference performance considerations.



\subsection{Justification}
\label{sec:confidence_threshold}

\begin{figure}[tb]
    \begin{minipage}[t]{\linewidth}
        \begin{subfigure}[t]{0.49\textwidth}
            \centering
            \includegraphics[width=\linewidth]{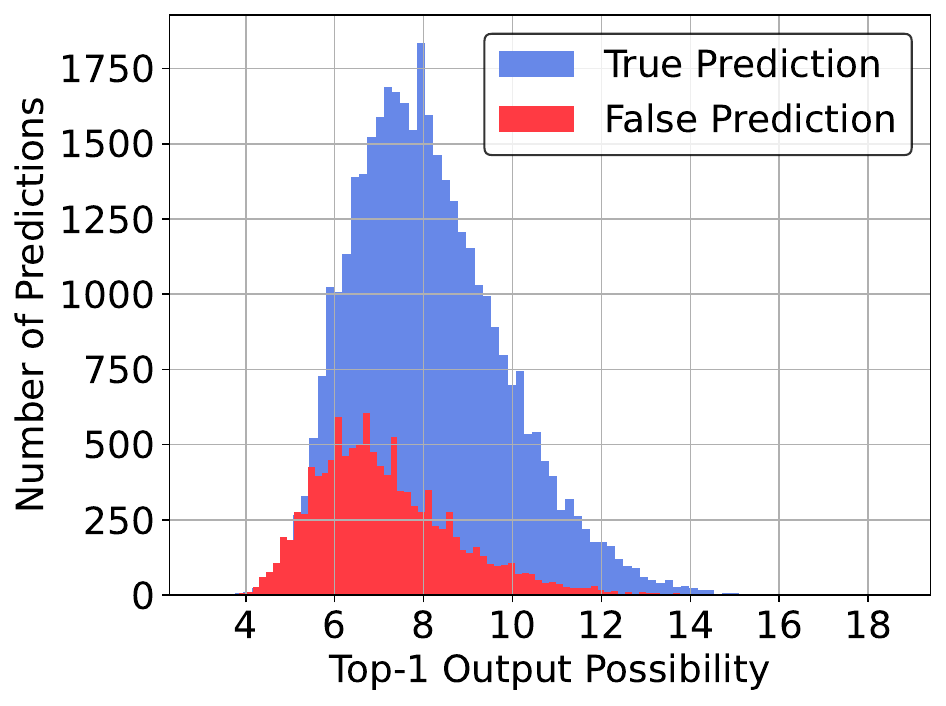}
        	\caption{Quantized MobileNet}
            \label{fig:boundary_fpga}
        \end{subfigure}
        \begin{subfigure}[t]{0.49\textwidth}
            \centering
            \includegraphics[width=\linewidth]{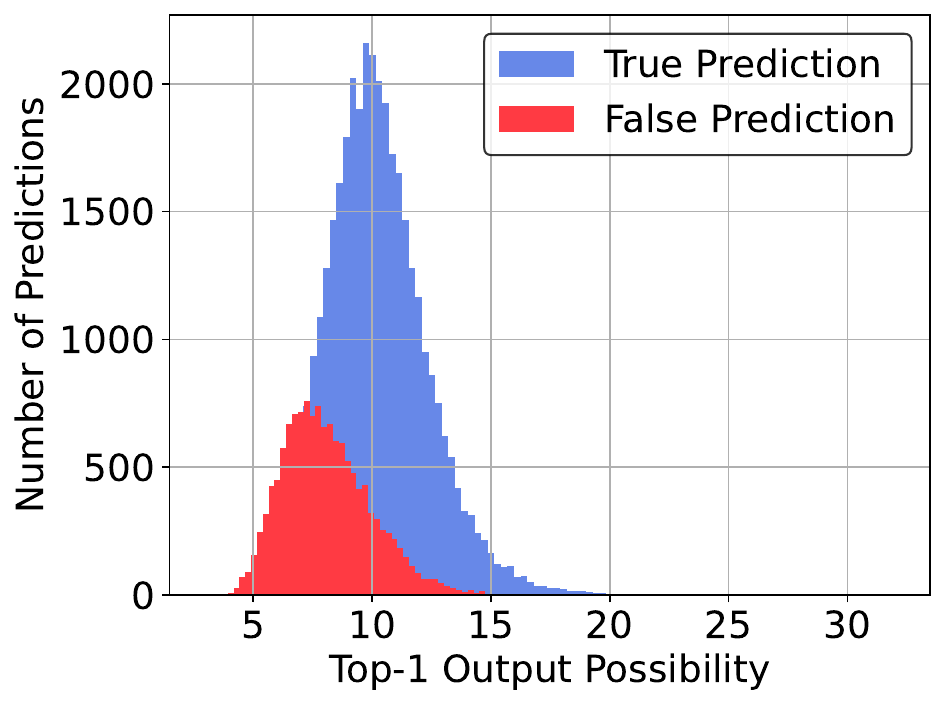}
        	\caption{MobileNet}
            \label{fig:boundary_gpu}
        \end{subfigure}
        \caption{Top-1 possibility distribution of true/false predictions of two models.}
        \label{fig:boundary}
	\end{minipage}
    \vspace{-8mm}
\end{figure}

Previous works~\cite{pmlr-v70-guo17a, Vemuri:EECS-2020-132,NEURIPS2021_8420d359} have shown that recent models are well-calibrated and the output possibility vector can correctly reflect the confidence of the inference results.

\begin{figure}[!tb]
    \begin{minipage}[t]{\linewidth}
        \begin{subfigure}[t]{0.49\textwidth}
            \centering
            \includegraphics[width=\linewidth]{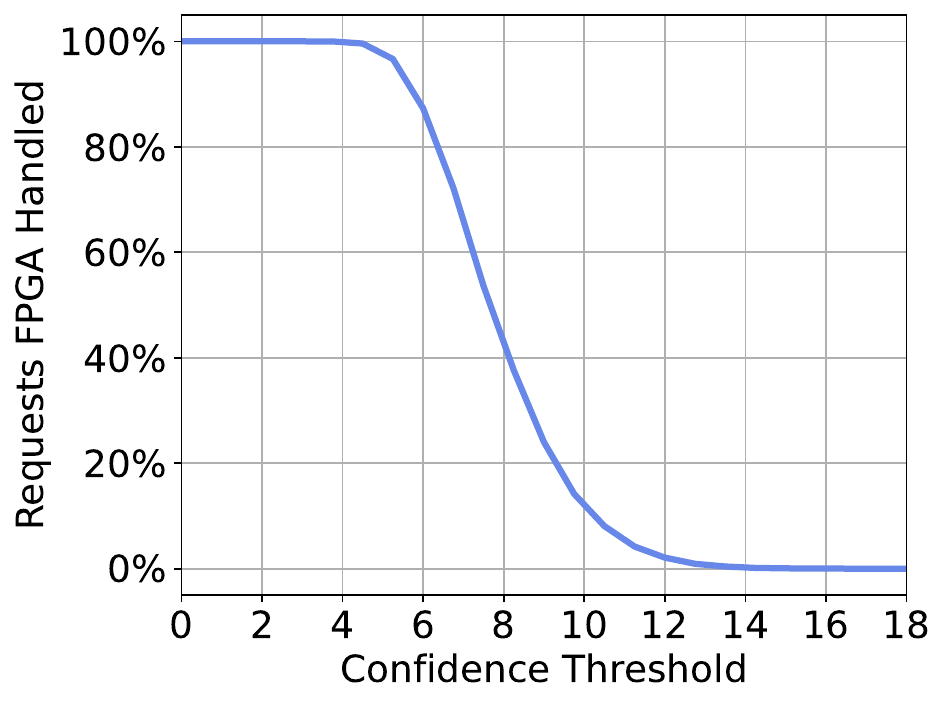}
        	\caption{FPGA handled}
            \label{fig:fpga_percent_score}
        \end{subfigure}
        \begin{subfigure}[t]{0.49\textwidth}
            \centering
            \includegraphics[width=\linewidth]{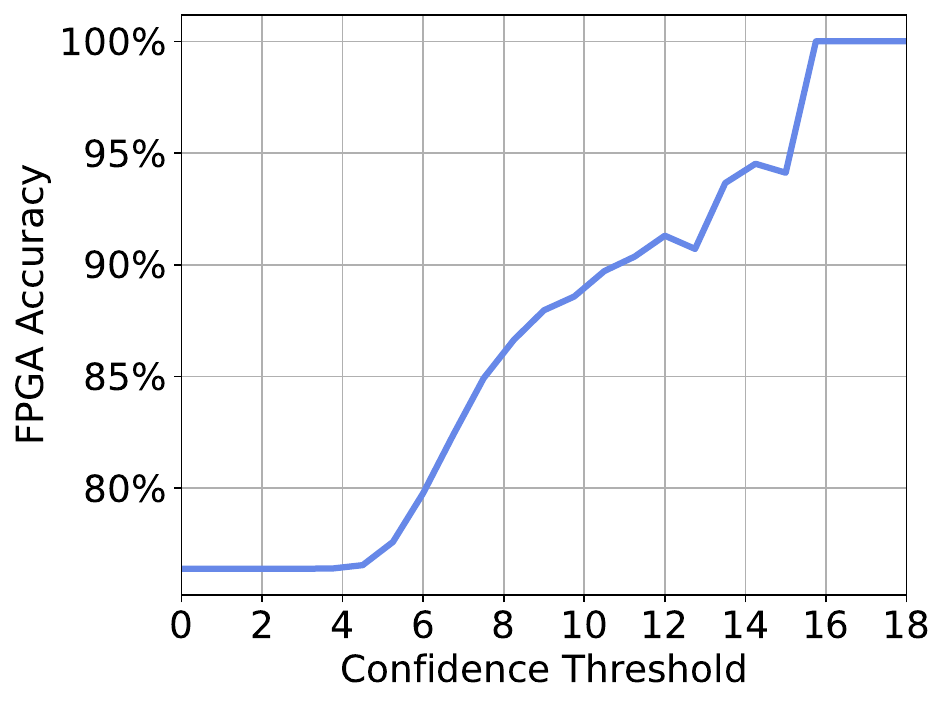}
        	\caption{FPGA Accuracy}
            \label{fig:fpga_accuracy_score}
        \end{subfigure}
        \caption{FPGA in different confidence threshold}
        \label{fig:fpga_accuracy_score_analysis}
	\end{minipage}
        \vspace{-3mm}
\end{figure}

To demonstrate this, we choose the MobileNet and its quantized version as examples and show the top-1 output possibility and the distribution of true and false predictions. 
The result is shown in Figure~\ref{fig:boundary}. 
In both models, true predictions have an overall higher confidence score than false predictions.
Next, we look at quantized MobileNet and show the effect of confidence thresholds.
We gradually increase the confidence threshold and accept the result only when the top-1 possibility is above the threshold.
Figure~\ref{fig:fpga_accuracy_score_analysis} shows the percentage of inference requests accepted and the corresponding accuracy.
When the confidence threshold increases, fewer requests are handled by the light model but the accuracy goes higher.



\begin{figure}[!tb]
    \begin{minipage}[t]{\linewidth}
        \begin{subfigure}[t]{0.49\textwidth}
            \centering
            \includegraphics[width=\linewidth]{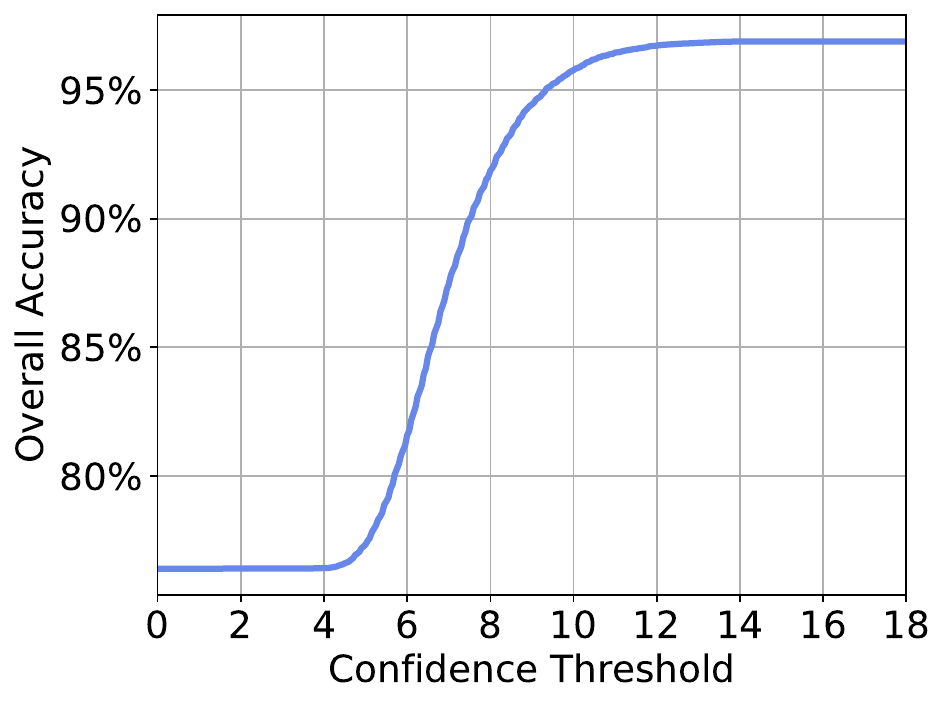}
        	\caption{Overall Accuracy}
            \label{fig:accuracy_with_score}
        \end{subfigure}
        \begin{subfigure}[t]{0.49\textwidth}
            \centering
            \includegraphics[width=\linewidth]{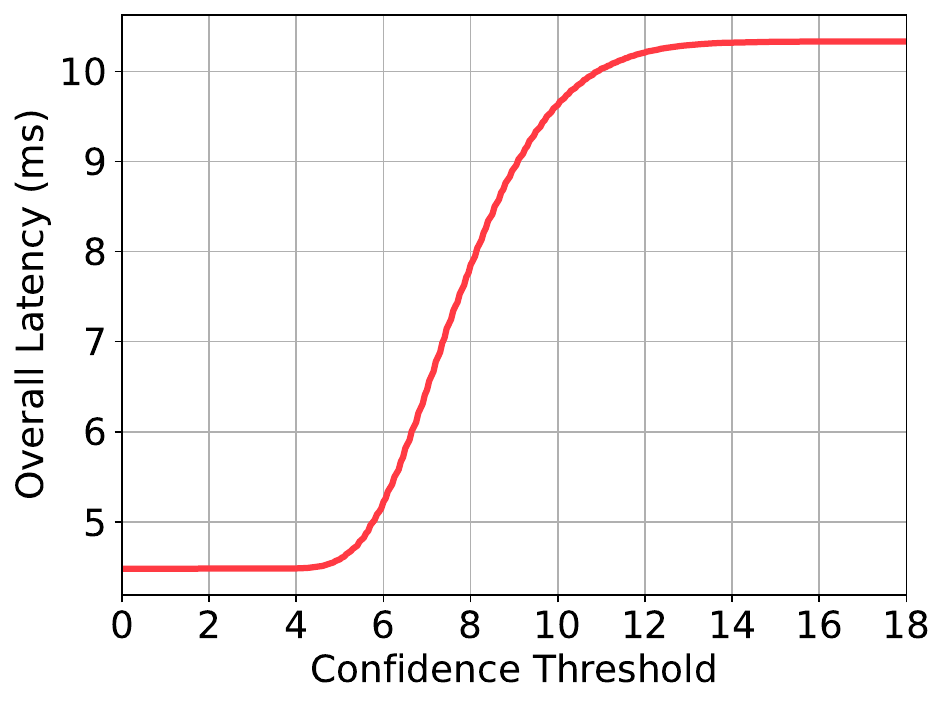}
        	\caption{Overall Latency}
            \label{fig:latency_with_score}
        \end{subfigure}
        \caption{Overall Accuracy v.s. Latency}
        \label{fig:confidence_score_tradeoff}
	\end{minipage}
        \vspace{-7mm}
\end{figure}

The confidence threshold offers a knob for the operator to choose the number of inference requests intercepted by the frontend, which 
balances between inference latency and accuracy: a lower confidence threshold offloads more work to the frontend, 
which lowers inference latency at the cost of lower accuracy, and vice versa. 
Figure~\ref{fig:confidence_score_tradeoff} illustrates the trade-off between latency and accuracy under different confidence thresholds. For instance, when the frontend processes only the top 50\% of high-score requests, the accuracy improves from 76\% to 85\%. Operators can adjust the confidence threshold to meet their target latency or accuracy goals based on application requirements.

\section{Lossy inference } 
\label{sec:unreliable_transport}

\sysname employs lossy inference to mitigate the impact of edge network fluctuations (e.g., reducing network retransmissions). Since a single lost packet can severely degrade inference accuracy—potentially corrupting the image and disrupting the ongoing task—the primary challenge is to safeguard the compressed image from packet loss while preserving accuracy as much as possible. Popular compression formats like PNG and JPEG add further complexity due to their reliance on two key design principles: \textit{(1) variable-length compression and (2) differential encoding}, both of which hinder recovery and complicate lossy inference.


We begin by showcasing our designs using JPEG~(\secref{sec:jpeg_format}), one of the most widely used encoding formats, to demonstrate how \sysname addresses \textit{variable-length compression} through basic unit protection. This design effectively limits the burst radius of packet losses, ensuring the image remains recoverable~(\secref{sec:basic_unit}).
Next, we explain how \sysname recovers lost information in images~(\secref{sec:delta_encoding}) to minimize the accuracy degradation of the inference model. Instead of filling zeros—which disrupt packet similarity due to \textit{differential encoding}—\sysname preserves accuracy by replacing missing data with already-received MCUs appended at the end of the image.




\subsection{JPEG as an Example}
\label{sec:jpeg_format}

One JPEG image has two sections: the header and the encoded payload.
The header stores the encoded image's metadata (such as dimension, customized application data, \emph{etc.})
and a collection of necessary information for the decoder.
The header is crucial to image decoding and the image is not recoverable if the header is damaged.
Since the header is relatively short compared with the encoded payload (usually less than 1\%), 
\sysname does not protect or recover the JPEG header if it is lost or partially lost.
Our focus is the encoded payload.

The basic unit of encoded payload is Minimum Coded Units (MCU). 
An MCU contains the compressed pixels of a square block (e.g., 8x8 pixels) in the original image.
Its length varies depending on the content in the block and it is not byte-aligned.
Usually, an MCU's size is tens of bytes, and the encoded payload contains hundreds to thousands of MCUs.
Each MCU is encoded based on the delta of the previous MCU.
Different MCUs are encoded independently, meaning the decoding pipeline is not disrupted if an entire MCU is missing. However, if part of the code within an MCU is lost, recovering the partially lost content becomes impossible.
To make things worse, the decoding pipeline can stall because the decoder cannot recognize the remaining code. This creates a significant "burst radius," as the decoder continues processing subsequent packets, consuming their content until it believes it has identified all MCUs specified in the header's metadata.  
Therefore, simply ignoring lost network packets without retransmission is not a viable option.

\subsection{Identify and Protect the Basic Unit}
\label{sec:basic_unit}
To prevent the image decoding pipeline from being interrupted by packet loss, we propose the basic-unit-aware packet packing (e.g., MCU in JPEG) on the \textit{variable-length code}.
The transport layer is aware of the sent image's encoding format and the sent packets are aligned with MCU as much as possible.
However, this is not easily achievable since a packet is always byte-aligned but an MCU is not.
To solve this problem, we propose a new byte-aligned layer atop the MCUs --- \textit{MCU Block}.
An MCU block contains multiple MCUs and always ends with a byte-aligned MCU. 
The MCU block is the basic recovery unit.
One MCU block may cross multiple packets, and if one of the packets is lost, the entire block is discarded.
We first present how \sysname constructs packets and the necessary information in the header to guide image decoding,
next, we explain how the image is recovered when a packet is lost.

\begin{figure*}[t]
\begin{minipage}[t]{0.4\textwidth}
    \centering
    \includegraphics[clip, trim=0cm 4cm 0cm 0cm,width=1\linewidth]{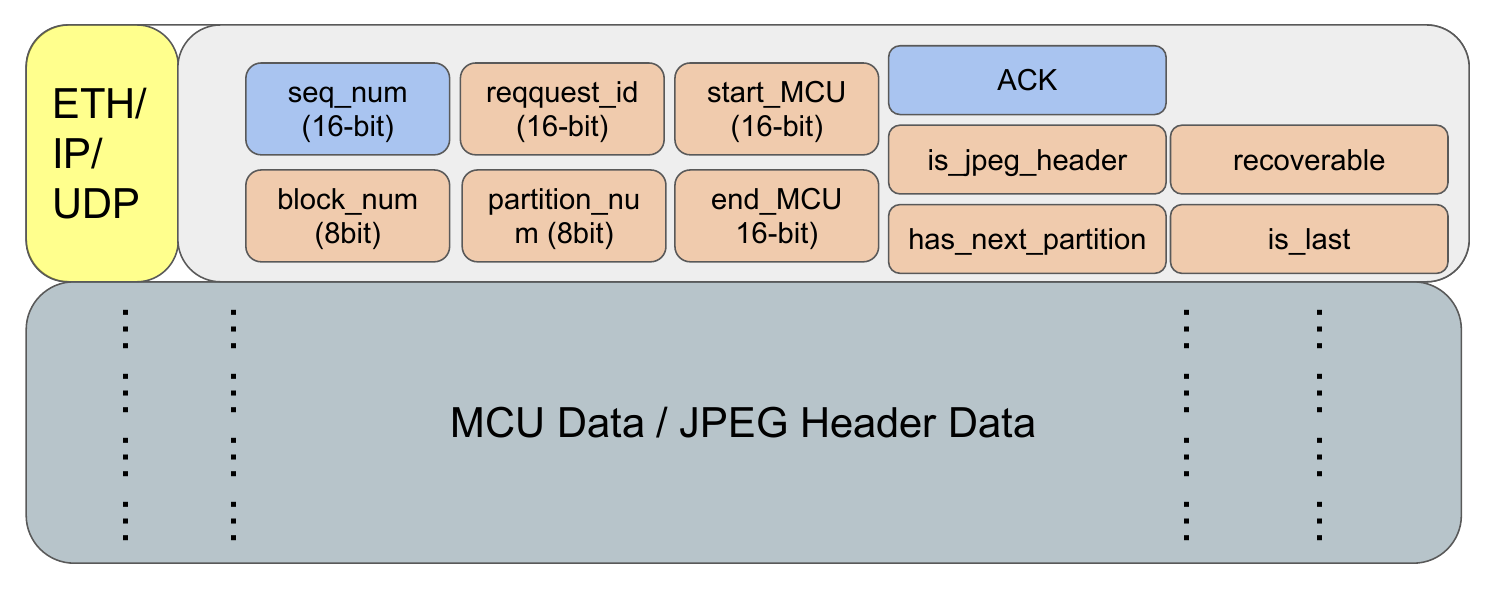}
	\caption{\sysname packet format, orange fields are for loss recovery protocol and blue fields are for reliability between clients and GPU servers.}
    \label{fig:packet_format}
\end{minipage}
\begin{minipage}[t]{0.29\textwidth}
\vspace{-8em}
    \begin{minipage}[t]{\linewidth}
        \begin{subfigure}[t]{0.49\textwidth}
            \centering
            \includegraphics[width=\linewidth]{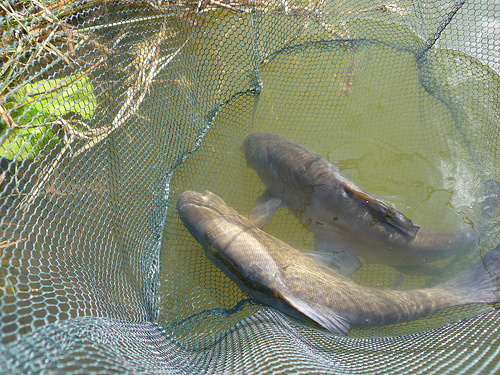}
        	\caption{Intact Image}
            \label{fig:loss0_visualize}
        \end{subfigure}
        \begin{subfigure}[t]{0.49\textwidth}
            \centering
            \includegraphics[width=\linewidth]{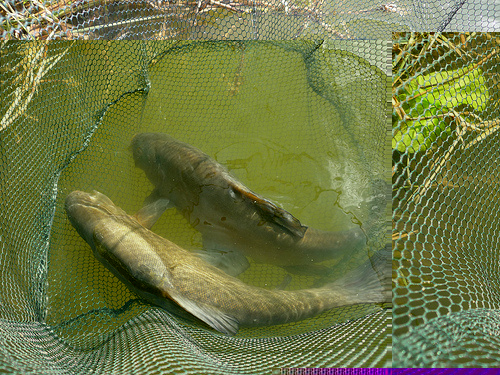}
        	\caption{Recovered Image}
            \label{fig:loss1_visualize}
        \end{subfigure}
        \caption{Original and recovered images} 
        \label{fig:loss_visualize}
	\end{minipage}
\end{minipage}
\begin{minipage}[t]{0.3\textwidth}
    \centering
    \includegraphics[width=0.7\linewidth]{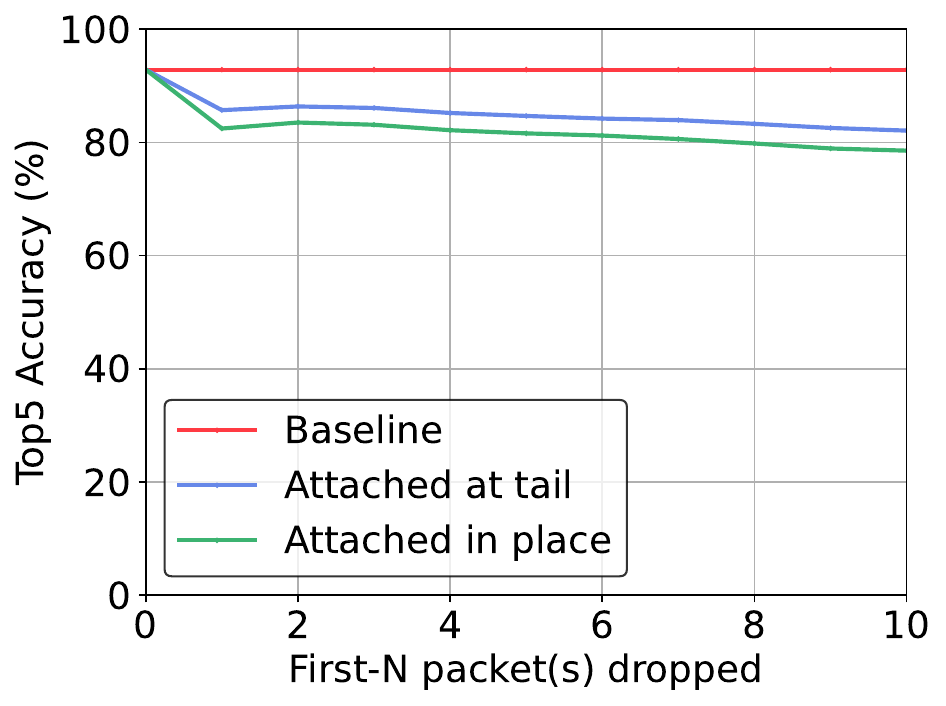}
	\caption{First-N-packets drop}
    \label{fig:partial_image_accuracy}
\end{minipage}
\vspace{-6mm}
\end{figure*}

\mypar{Packets Construction.}
\label{sec:packet_construction}
The guideline for designing the header format and packet construction process is to provide as much 
information as possible to simplify and accelerate the processing pipeline.
This helps avoid additional processing in the frontend, which stalls the pipeline and increases the inference latency.

Before sending the packet, the sender examines the image according to the JPEG format and splits the image into header and multiple MCU blocks.
Next, \sysname constructs the sent packets by packing each MCU block into as few packets as possible,
then set the following header fields, as shown in Figure~\ref{fig:packet_format}:

\begin{itemize}[itemsep=0pt, parsep=0pt, topsep=0pt, partopsep=0pt]
    \item \textit{request\_id.} The unique ID of each inference request.
    \item \textit{start\_MCU} and \textit{end\_MCU.} The start and end MCU IDs in the MCU block. This helps the frontend to recognize if MCUs are missing and how many are missed so that they can be recovered accordingly.
    \item \textit{block\_num.} Indicates which block the packet belongs to, \emph{i.e.} whether the packet belongs to the current MCU block or the next one.
    \item \textit{partition\_num.} Indicates how many packets the MCU block is split into. This tells the receiver how many packets should be expected since frontend has to drop the entire MCU block if any packet is missing.
    \item \textit{has\_next\_partition.} This 1-bit field denotes whether the next packet belongs to the same MCU block.
    \item \textit{is\_jpeg\_header.} If the current packet contains the JPEG header, this 1-bit field is set. 
    Because the JPEG header is not recoverable, if lost, the frontend skips the current image, clears the cache, and lets the backend handle the inference.
    \item \textit{is\_last.} This indicates the last packet of the image and triggers the image recovery pipeline if any packet is lost during transmission.
    \item \textit{recoverable.} If the image is too small, \sysname cannot find an MCU block or there is only one such block.
    This bit is used to notify the frontend to flush the entire image when any packet is lost because such an image is not recoverable.
\end{itemize}

Upon receiving the first packet of the image, the frontend examines whether the \textit{is\_jpeg\_header} field is set, 
if not, the header is lost during transmission and the image is not recoverable, all subsequent packets with the same 
\textit{request\_id} are ignored. 
Otherwise, the frontend proceeds the decoding and recovery procedure.
The frontend stitches packets together into MCU blocks by examining the \textit{has\_next\_partition}, 
\textit{block\_num}, and \textit{partition\_num} fields. 
If any packets are lost, the frontend records how many MCUs are lost and discards the entire MCU block.

\vspace{-1mm}
\subsection{Delta Encoding Resilience}
\label{sec:delta_encoding}


As introduced in Section~\secref{sec:jpeg_format}, our recovery design aims to restore the image after packet loss while preserving as much accuracy as possible.
A naive recovery approach is to insert the recovery MCU directly into the stream upon detecting a missing MCU block (packet loss). However, MCU encoding relies on \textit{differential encoding}, where each MCU depends on its preceding MCU for encoding differences. Randomly selecting a recovery MCU from the image (or filling with a black MCU), can introduce incorrect information, propagating errors throughout the rest of the image due to \textit{differential encoding}, thereby reducing the accuracy of the inference model.

Instead, \sysname bridges the gap between the existing MCU and the incoming MCU directly when MCU loss (packet loss) occurs and places the recovery MCU (or a black MCU) at the end of the image. This approach leverages the spatial similarity between nearby MCU blocks, minimizing the impact of the lost information. Figure~\ref{fig:loss_visualize} shows the original image alongside the recovered one after a packet is lost during transmission. The objects in the image remain intact, and the decoder at the frontend continues to produce the entire image.

When some MCU blocks are lost, the images may exhibit line shifting and color distortion due to the missing MCUs but largely retain complete and identifiable components, thanks to spatial similarity and differential encoding. This property is not limited to specific datasets or models. Furthermore, modern vision models, which heavily rely on convolutional techniques, are inherently tolerant to missing information.

To further study on this, we set up an inference service running the ResNet50 model and tested it with the entire ImageNet dataset.
We dropped the first N packets for every image during transmission under different loss rates and recovered it through the two above recovery procedures. 
The first N packets impact the images most because all the following MCUs' delta values are changed. 

The average image size of the ImageNet is 130k bytes. The result shows even if 10 packets are dropped (around 13k bytes, nearly 10\%), the model remains around 80\% of top-5 accuracy. 
Attaching the lost MCU at the tail of the image gives another 4\% accuracy boost compared with attaching them in place.

While we showcase JPEG, our approach generalizes to other compression formats too. For PNG, the basic unit could be the IDAT chunk. Non-compressed formats like BMP and RAW can also be treated as a special case within our framework.
\section{GPU-based System Implementation}\label{sec:impl} 
\label{sec:implementation}

The \sysname implementation consists of (i) a client that launches inference requests, (ii) a frontend GPU-based edge inference box, (iii) a backend GPU-based data center inference server.


\mypar{Client.}  
The client launches prediction requests. We develop a communication library handles the image input from the users. The library has two main functions: (1) parsing and collecting the necessary metadata from the request image to support our loss-tolerant protocol. We use libjpeg-turbo~\cite{libjpeg-turbo} to analyze the image and collect metadata. (2) Constructing the \sysname protocol packet format with the metadata before sending the packets. This component is built entirely in C++ using DPDK to ensure high performance.

\mypar{Frontend.}  
The \sysname frontend is a modelless inference serving system built in C++ that uses TVM~\cite{tvm} as its inference engine, offering flexible control over model loading and unloading. To enable lossy inference and recover images from corruption in the JPEG format, we use the same communication library to handle our customized network protocol. This is complemented by a high-performance JPEG recovery module that implements the loss recovery mechanism (Section~\secref{sec:unreliable_transport}) before forwarding the request to the inference engine. Section~\secref{sec:implementation_fpga} describes the FPGA frontend alternative, which provides a more cost-effective hardware implementation.


\mypar{Backend.}
We built a proxy for the backend that can integrate with any serving systems, such as Clockwork, InFaaS, and NVIDIA's Triton Inference Server~\cite{triton}. The proxy manages requests using a reliable protocol and includes a dedicated channel for processing notifications from the frontend. Upon receiving these notifications, it coordinates GPU interruptions or removes requests from the GPU queue. To ensure a valid baseline evaluation, we adapted Clockwork to support modelless inference, with comparisons provided in Section~\secref{sec:eval}.


\section{FPGA frontend Prototype Implementation}
\label{sec:implementation_fpga}

\begin{figure}[t]
    \centering
    \includegraphics[clip, trim=0cm 0cm 0cm 0cm, width=\linewidth]{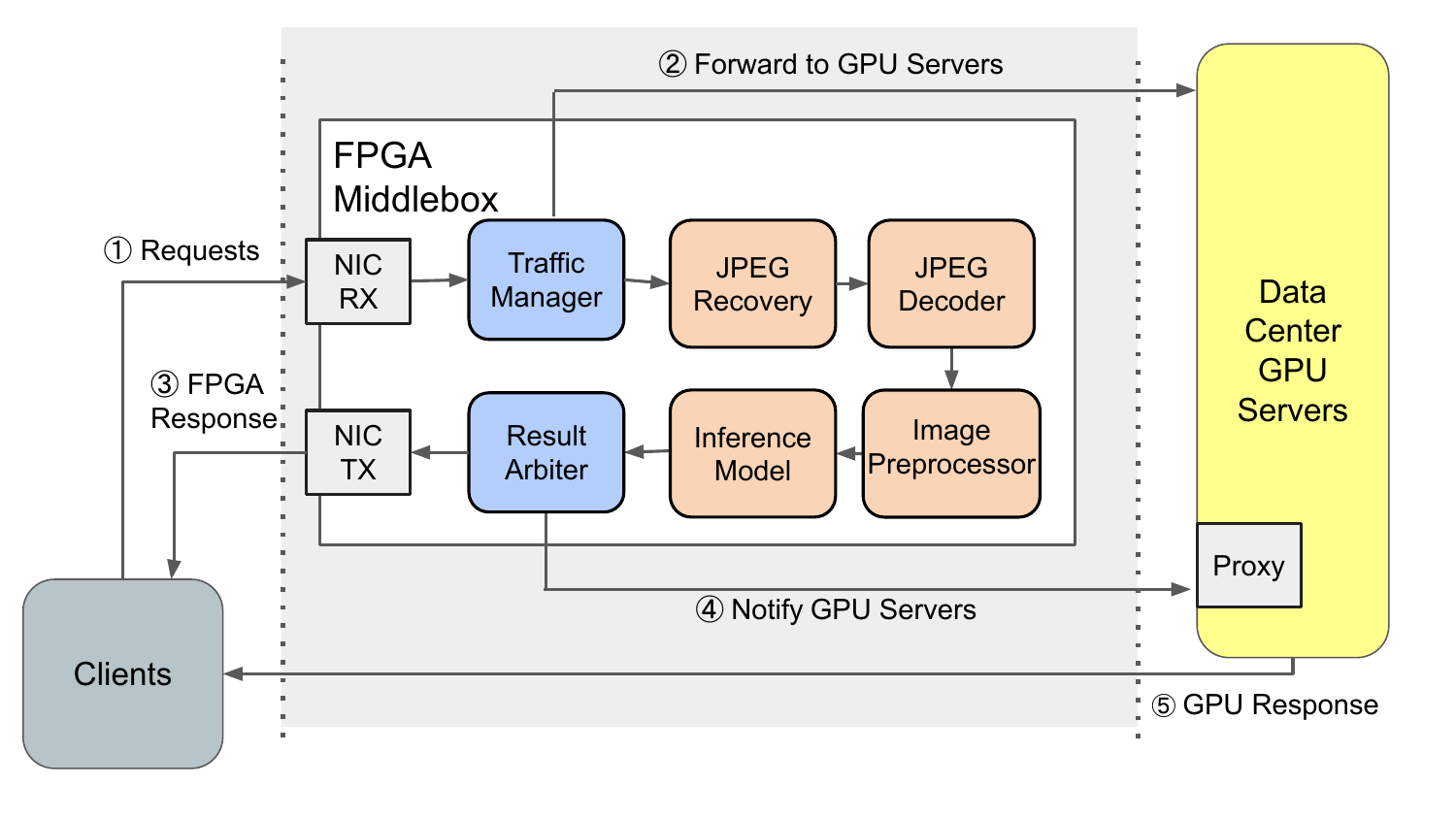}
	\caption{\sysname-FPGA Implementation Overview}
    \label{fig:nic_overview_fpga}
    \vspace{-6mm}
\end{figure}



Power consumption is important at the edge. We also implemented an FPGA frontend prototype to demonstrate the power consumption and performance benefits by implementing the entire inference pipeline to support modeless serving, and to validate our hardware-friendly designs, \textit{confidence scaling} and \textit{lossy compression recovery}, as shown in Figure~\ref{fig:nic_overview_fpga}.

We use the Xilinx Alveo U250 FPGA~\cite{u250} as our development board. The board is equipped with 2x QSFP28 network interfaces. The inference model complies with the modified FPGA AI model compiler FINN~\cite{blott2018finn, finn} so that the model can output both answers and probability vectors. The FPGA JPEG decoder is sourced from an open-source solution~\cite{core_jpeg}. For other components, we write P4~\cite{p4}, C++ (Vitis HLS~\cite{vitishls}), and Verilog for different purposes to generate the final RTL code. The Vivado~\cite{vivado} and Vitis HLS versions used are 2021.2. 

The only functional difference between the FPGA prototype and the GPU-based frontend is the support for swapping. We did not implement swapping in the FPGA. Consequently, the prototype supports only one application at a time (one small model in the FPGA frontend and one large model in the GPU backend). 

The FPGA frontend processes packets sequentially through a series of pipeline modules, including the Traffic Manager~(\secref{fpga:traffic_manager}), Image Recovery Module~(\secref{fpga:recovery}), Image Decode and Preprocessor~(\secref{fpga:image}), Inference Engine~(\secref{fpga:inference_model}), and Result Arbiter~(\secref{fpga:arbiter}). The implementation details of the FPGA prototype are described below.



\subsection{Traffic manager}
\label{fpga:traffic_manager}
The Traffic Manager serves as the first entry point for request packets arriving at the FPGA's CMAC module (Rx). It processes incoming packets by duplicating them for both frontend and backend operations. The original packet is forwarded to the next component, while the duplicate is sent to the GPU. This design ensures the seamless and efficient operation of our two-layer architecture.

In addition, the Traffic Manager handles retransmitted packets by forwarding them directly to the backend, as the FPGA is designed for lossy inference and does not manage retransmissions. This approach simplifies hardware design by avoiding the need to implement a complex reliability protocol in the FPGA. Instead, the backend employs a reliable protocol such as TCP. The Traffic Manager also filters out packets unrelated to \sysname.

\subsection{Image Recovery Module} 
\label{fpga:recovery}
The packets then stream to the Image Recovery Module. The recovery module restores JPEG images when packet loss occurs, ensuring a decompressible format for the subsequent JPEG decoding component. To maintain a continuous streaming process, we do not wait for or handle lost packets; instead, we bridge the existing MCU with the MCUs from the next packet as they arrive. However, since MCUs have variable lengths and operate at the bit level, the bridging process requires precise bit-level alignment of the JPEG code to seamlessly connect two MCUs that were originally unaligned. Any bit error during the decompression process can stall the entire operation, making accurate alignment and handling critical.

To address this, we add a buffer line in the FPGA to temporarily store the last line (each line corresponds to the data width processed in a single cycle by the FPGA) of the latest MCU, rather than streaming the code directly to the next component. When the next MCU is a bridged MCU, all bits must be offset to seamlessly connect it with the existing MCU. This process requires shifting all subsequent bytes in the following packets by the same offset.

We also maintain a count of lost MCUs. Once the final packet is received or a timeout occurs, the recovery module uses the count of lost MCUs to append a series of recovery MCUs (or MCUs with black pixels) to the end of the JPEG code. This step ensures decoding correctness, as the number of MCUs decoded is defined in each JPEG. A mismatch in the MCU count during decoding can potentially stall the process and propagate incorrect results to subsequent images. This approach enables the FPGA to continue inference without discarding the image due to packet loss. The entire component is implemented using Vitis HLS.

\subsection{Image Decoder and Image Preprocessor} 
\label{fpga:image}
The Image Decoder module converts the JPEG byte stream into a raw RGB data stream, which is then passed to the subsequent Image Preprocessor. The Preprocessor processes various types of RGB data stream into a format suitable for input to the inference model.

Our JPEG decoder is modified from an open-source implementation~\cite{core_jpeg} with several bug fixes. It is capable of handling various versions of JPEG code and can parse any resolution found in the ImageNet dataset, making it a general-purpose JPEG decoder. The decoder operates without any assumptions about the data, as long as it is valid JPEG code. This module can be replaced with any commercial JPEG decoder if needed. 

After decoding, the output is a raw RGB data stream. However, since image metadata, such as dimensions (e.g., width and height), vary during encoding and cannot be controlled, preprocessing is required to adjust the input before feeding the data into the model. Following the practices outlined in the ImageNet training script~\cite{imagenettrainingscript}, this module performs image resizing and cropping as part of the preprocessing.
Since image processing can be streamlined and is highly parallelizable (hardware-friendly), the module delivers high performance compared to CPU-based solutions and achieves performance comparable to the state-of-the-art GPU image processing library, DALI~\cite{dali}, while consuming less power.

We have decoupled this component as a separated module to support different image processing requirements. It is written in C++ using Vitis HLS, allowing developers to modify this module independently to accommodate specific image processing needs.

\subsection{Inference Engine} 
\label{fpga:inference_model}
The FPGA employs a simpler yet faster model to accelerate inference tasks. To maintain good overall accuracy, the model outputs the inference result along with the corresponding confidence score, which determines whether the results should be returned directly or ignored.

Although many academic works propose models for FPGA deployment (as discussed in Section~\secref{dis:fpga}), development on FPGAs often lags behind the rapid advancements in machine learning, making FPGAs challenging to deploy. By leveraging confidence scores, FPGA only needs to expose confidence scores to the backend GPU or handle simpler requests independently. This two-layer design minimizes the need for frequent model updates on the FPGA, enhancing its practicality and enabling seamless collaboration with the GPU.

We modified the MobileNet model developed with Xilinx FINN. The original model outputs only the top-5 results. Our modifications enable the model to output both the top-5 results and the softmax values (representing confidence scores) in the wire-level. The confidence scores and top-5 results are streamed to the next component, the arbiter, which determines whether to provide a direct response or forward the request to the backend based on a confidence threshold.


\subsection{Result Arbiter}
\label{fpga:arbiter}
The arbiter maintains a confidence score threshold and decides whether to respond to the user or ignore the results. It operates in two cases:

(1) Confidence above the threshold: If a result's confidence score exceeds the threshold (indicating a high likelihood of correctness), the arbiter streams the result to the FPGA CMAC module (Tx) to send it directly to the client. Simultaneously, it generates a terminate packet for the backend GPU to cancel the corresponding inference request. Both packets are sent using UDP and require no acknowledgment. As the entire design is implemented on the FPGA, there is no need for a separate CPU-based system. This design allows the FPGA to operate independently, optimizing performance and reducing power consumption.

(2) Confidence below the threshold: If a result's confidence score is below the threshold, the arbiter discards the result and lets the GPU complete the inference. The arbiter simply skips the result and begins processing another request. The backend GPU processes the request and forwards the result to the client.

\ifdefined\ArxivVersion
\else
    Full details on the implementation for the FPGA in each component can be found in our open-source repository~\cite{bifimpl}.
\fi

\section{Evaluation}\label{sec:eval}
\sysname demonstrates improvement with up to $1.6\times$ P90 improvement compared to the baseline for serving applications with three different GPU and multi-tenancy settings. Our FPGA prototype improves latency and power consumption by up to $1.5\times$ and $3.34\times$, respectively, compared to the A100. We also demonstrate the feasibility of our FPGA prototype performing lossy inference under severe packet loss. Additionally, we report the implementation resource details of \sysname with FPGA.

\subsection{Experimental Setup}
We evaluate \sysname in an edge-DC setting. To replicate edge-internet settings, we use the Linux \textit{tc} tool to add the edge delay from the client to the frontend and the internet delay to the backend, simulating real cases similar to previous work~\cite{AWStream_sigcomm18}. Our default setting uses 3ms and 10ms as edge and internet delays, respectively.
For both the frontend and backend, we use NVIDIA A100 GPUs but limit the memory usage to a range from 4GB to 16GB to reflect typical settings for edge hardware accelerators~\cite{gemel, jetson}.


We target the image classification task using the ImageNet~\cite{imagenet} dataset, which includes around 50K images with a wide range of resolutions, from as small as tens of pixels by tens of pixels to as large as 4K. We use a uniform random distribution as the workload distribution for users to send different accuracy requirements.

For the FPGA prototype (\sysname-FPGA), we built our system on an AMD Xilinx U250 FPGA to demonstrate the performance and power benefits and show the feasibility of hardware-friendly packet loss recovery and confidence scaling implemented in FPGA.




\mypar{Baselines.}
We pick Clockwork~\cite{clockwork} as our main baseline. Clockwork is a state-of-the-art inference serving system that schedules incoming user requests to different GPU instances with the goal of maintaining predictable performance. We apply Clockwork in the edge-DC setting by scheduling requests on two GPUs, one as the edge and one as the DC GPU. Since Clockwork targets a homogeneous GPU setting, we set the edge GPU and the DC GPU to have the same GPU memory size to make a fair comparison.

For the FPGA implementation, we use both T4 and A100 GPUs for comparison, to show the performance and power consumption with different accelerator options.




\mypar{Metrics.}
We report (1) the average latency, as well as latencies at P90, P99, and P100 for all inferences in the system. Tail latency, crucial for meeting Service Level Objective (SLO) requirements in inference serving~\cite{clockwork}, is one of our main motivations for optimizing latency. (2) Power consumption, calculated as the number of requests multiplied by the average power consumption per request for each device (FPGA/GPU/both), is crucial for edge settings with limited power resources. We compare the power efficiency across hardware accelerators and also report the FPGA resources used in our implemented prototype.

\ifdefined\ArxivVersion
    \begin{figure*}[t]
        \centering
        \begin{minipage}[t]{0.8\linewidth}
            \begin{subfigure}[t]{0.33\textwidth}
                \centering
                \includegraphics[width=\linewidth]{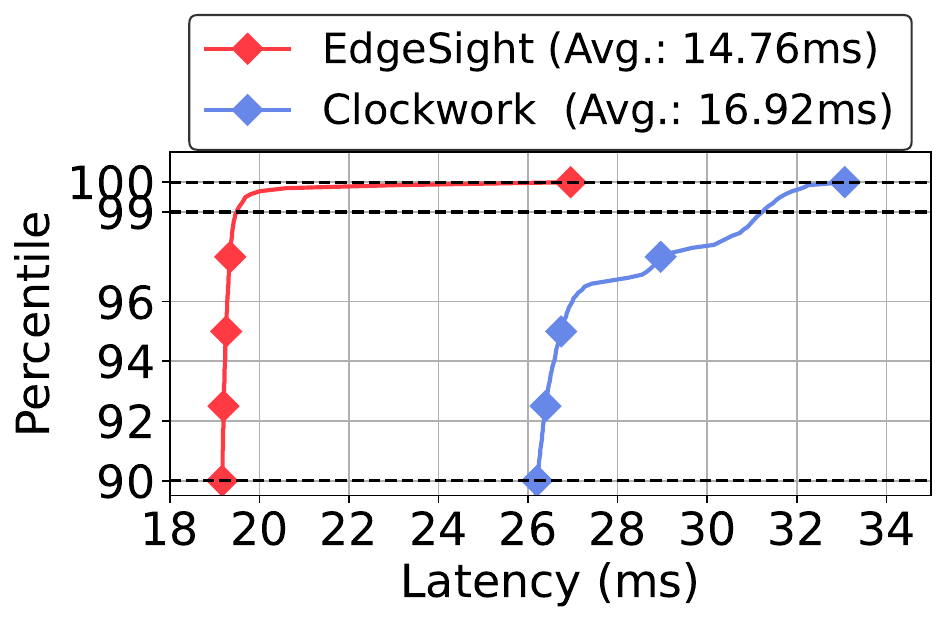}
            	\caption{4GB, 25 applications}
                \label{fig:4_25_concurrent}
            \end{subfigure}
            \begin{subfigure}[t]{0.33\textwidth}
                \centering
                \includegraphics[width=\linewidth]{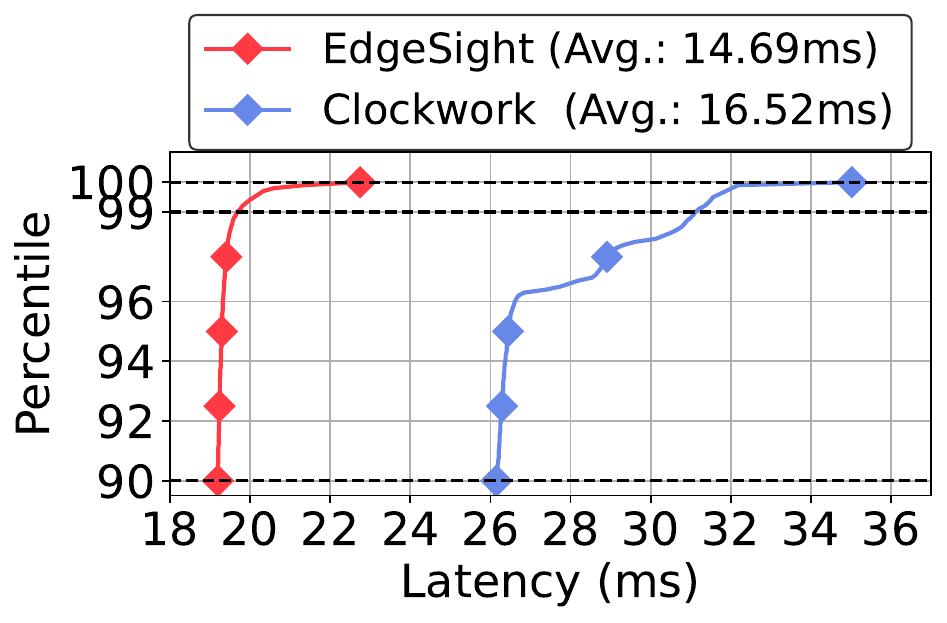}
            	\caption{8GB, 50 applications}
                \label{fig:8_50_concurrent}
            \end{subfigure}
            \begin{subfigure}[t]{0.32\textwidth}
                \centering
                \includegraphics[width=\linewidth]{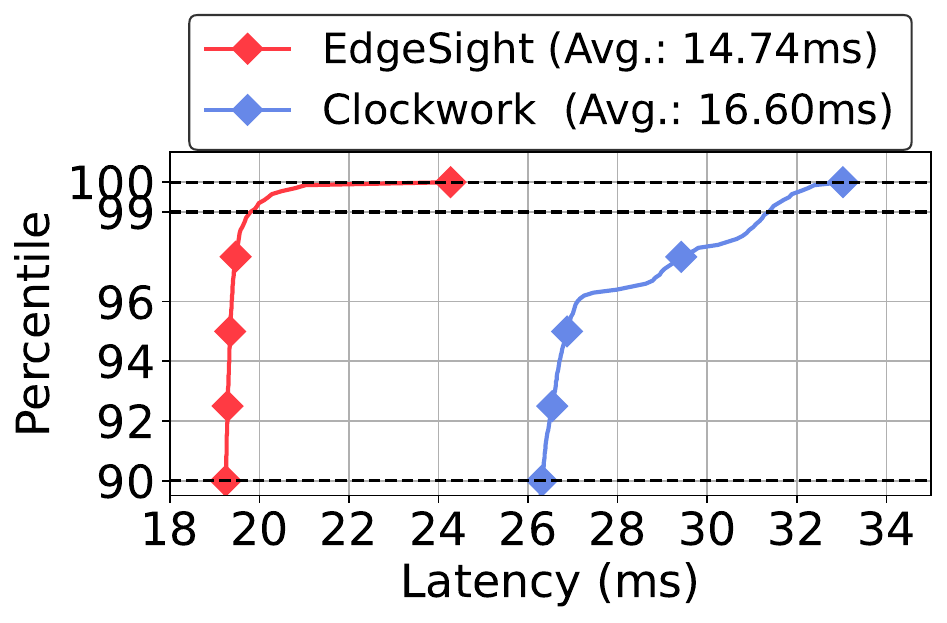}
            	\caption{16GB, 100 applications}
                \label{fig:16_100_concurrent}
            \end{subfigure}
            \caption{Latency percentiles of three settings}
            \label{fig:concurrent_swapping}
    	\end{minipage}
        \vspace{-2mm}
    \end{figure*}
\else
    \begin{figure*}[t]
        \centering
        \begin{minipage}[t]{0.8\linewidth}
            \begin{subfigure}[t]{0.33\textwidth}
                \centering
                \includegraphics[width=\linewidth]{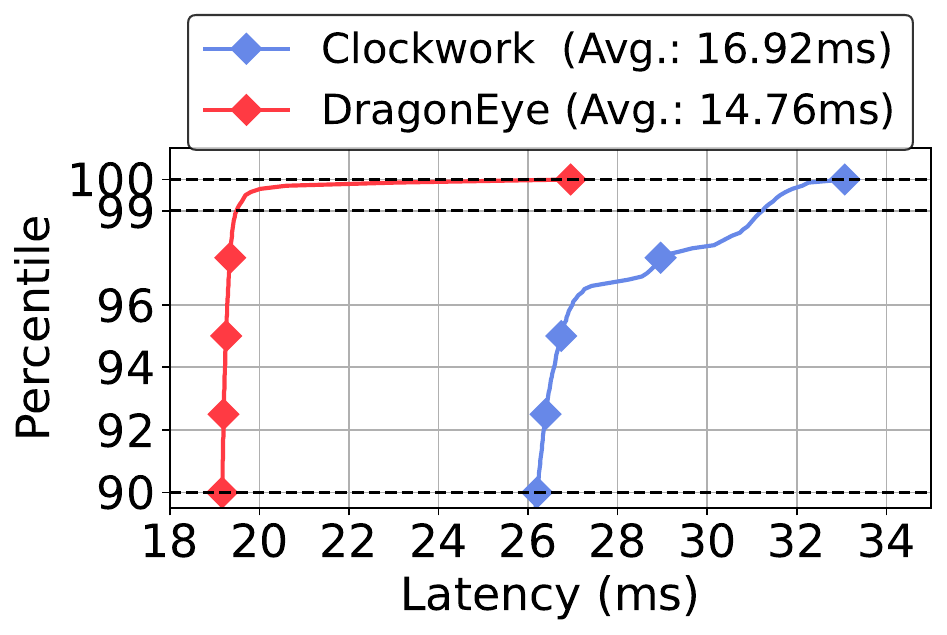}
            	\caption{4GB, 25 applications}
                \label{fig:4_25_concurrent}
            \end{subfigure}
            \begin{subfigure}[t]{0.33\textwidth}
                \centering
                \includegraphics[width=\linewidth]{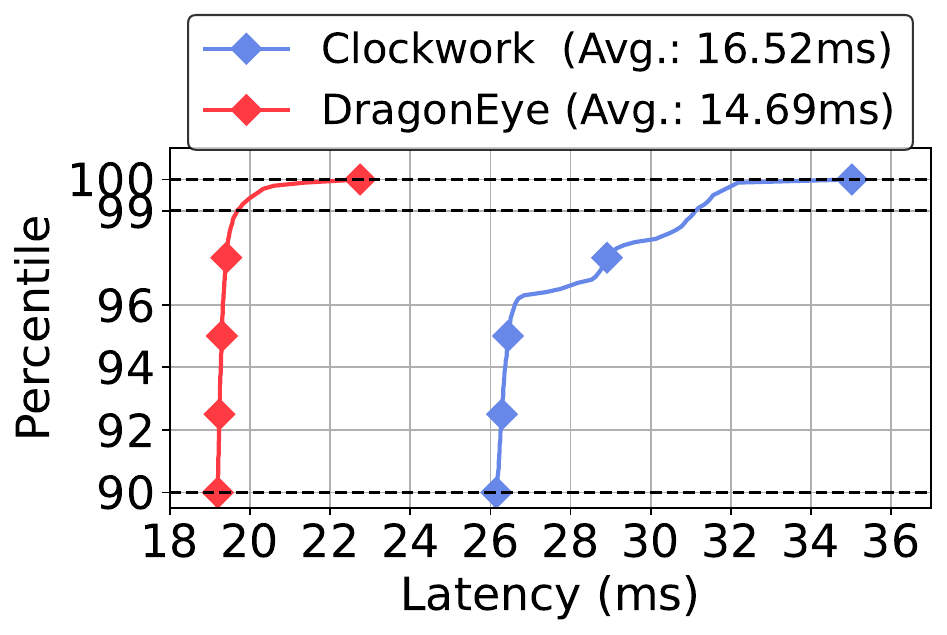}
            	\caption{8GB, 50 applications}
                \label{fig:8_50_concurrent}
            \end{subfigure}
            \begin{subfigure}[t]{0.32\textwidth}
                \centering
                \includegraphics[width=\linewidth]{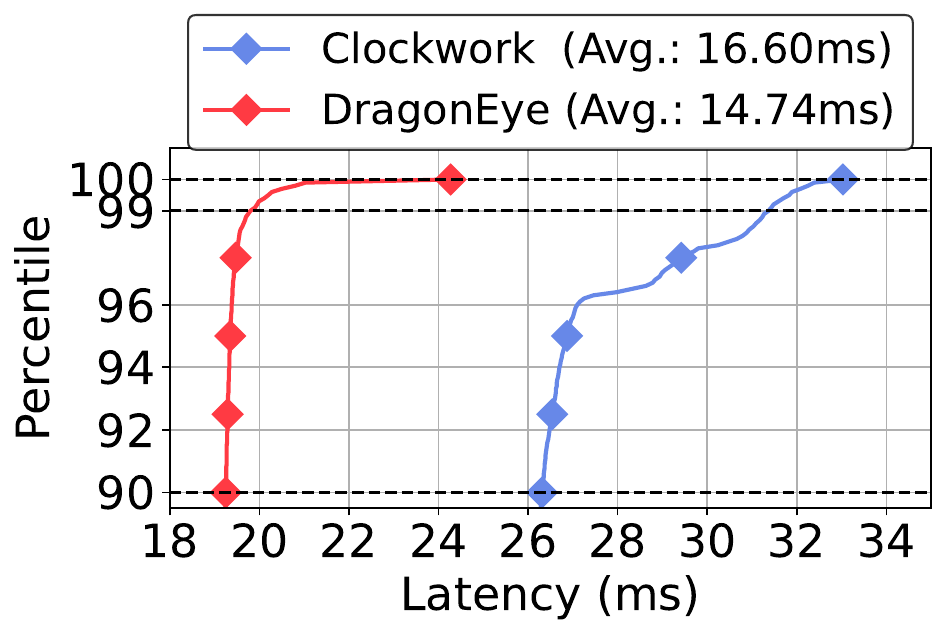}
            	\caption{16GB, 100 applications}
                \label{fig:16_100_concurrent}
            \end{subfigure}
            \caption{Latency percentiles of three settings}
            \label{fig:concurrent_swapping}
    	\end{minipage}
        \vspace{-2mm}
    \end{figure*}
\fi

\subsection{\sysname Reduces Swapping Cost}
\label{eval:edge}
We compare the edge-DC systems with \sysname and the traditional model separation scheme in Clockwork. We configure three different settings in terms of the number of applications and GPU memory for the edge-DC systems (2 GPUs) to serve: (4GB, 25 applications), (8GB, 50 applications), and (16GB, 100 applications). For each application, we use 5 models with different accuracy profiles to meet various accuracy requirements from different users, ranging from 70\% to 85\%, as aligned with \cite{Cocktail}.

Figure~\ref{fig:concurrent_swapping} shows the latency performance for these 3 settings. Taking Figure~\ref{fig:4_25_concurrent} as an example, 25 applications each with 5 models (totaling 25 * 5 = 125 models) will participate in the inference. When the system receive the request, the corrsponding models will first be loaded into the memory if they do not already exist, thereby incurring swapping latency. \sysname reduces the number of models from 125 to 50 because each application needs only two models to cover the accuracy requirement spectrum.

In the 25 application settings (Figure \ref{fig:4_25_concurrent}), \sysname achieves both better tail latency and average latency compared to Clockwork, outperforming it by $1.36X$, $1.6X$, and $1.22X$ in P90, P99, and P100, respectively. \sysname also surpasses Clockwork by $1.14X$ in average latency. The trends observed in the 50 (Figure\ref{fig:8_50_concurrent}) and 100 (Figure~\ref{fig:16_100_concurrent}) application settings are similar.

\ifdefined\ArxivVersion
    \begin{figure*}[t]
        \begin{minipage}{0.25\textwidth} 
            \centering
            \includegraphics[width=\linewidth]{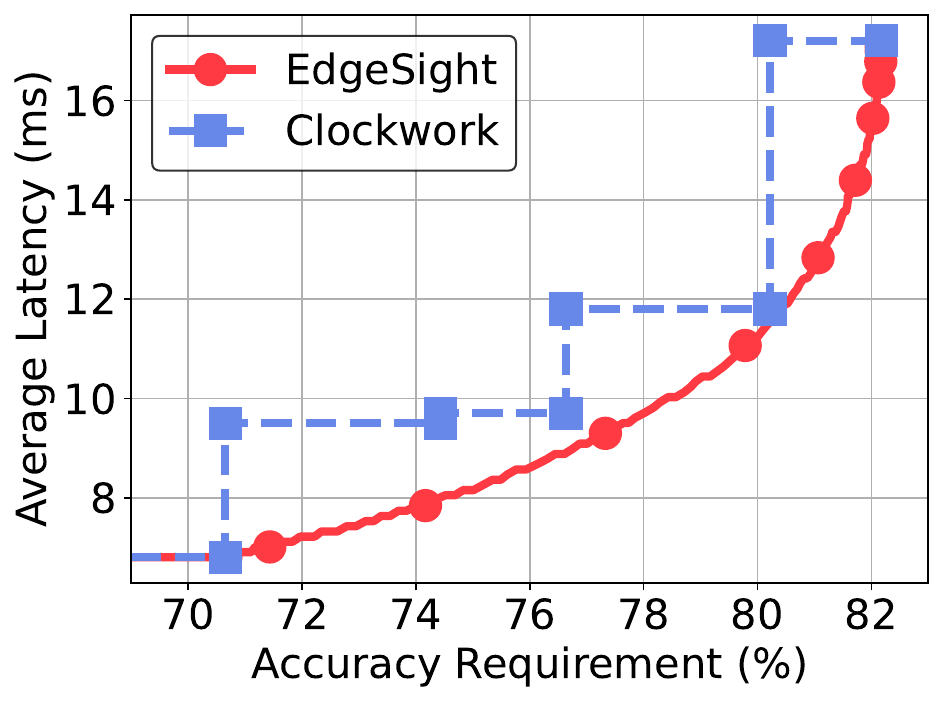}
        	\caption{Latency curve with different accuracy requirement}
            \label{fig:latency_curve}
        \end{minipage}
        \begin{minipage}{0.73\textwidth}
            \begin{subfigure}[t]{0.32\textwidth}
                \centering
                \includegraphics[width=\linewidth]{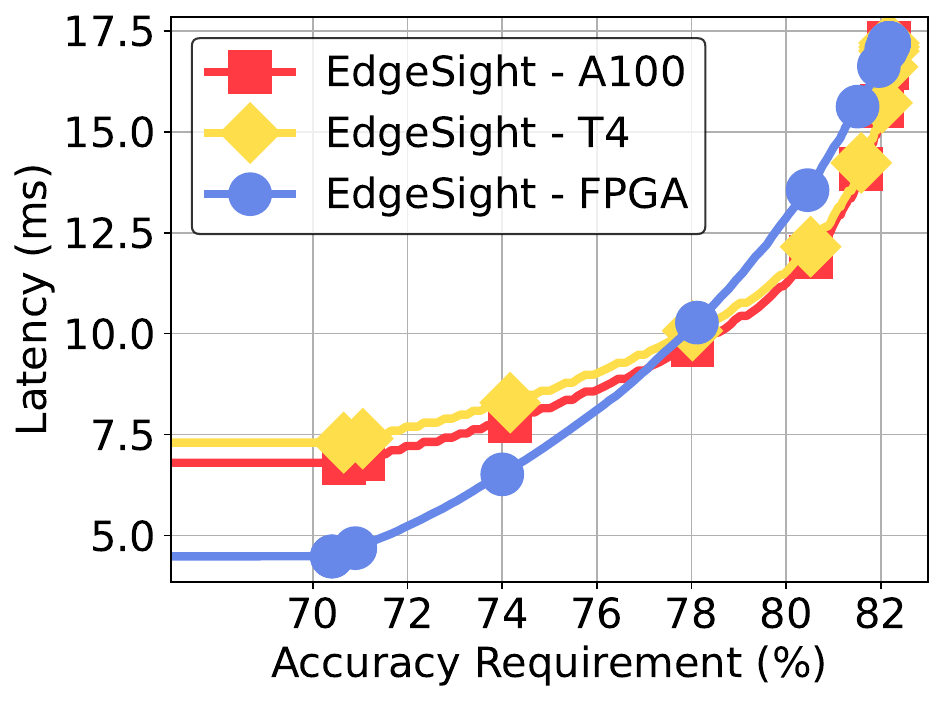}
            	\caption{Latency}
                \label{fig:fpga-performance}
            \end{subfigure}
            \begin{subfigure}[t]{0.32\textwidth}
                \centering
                \includegraphics[width=\linewidth]{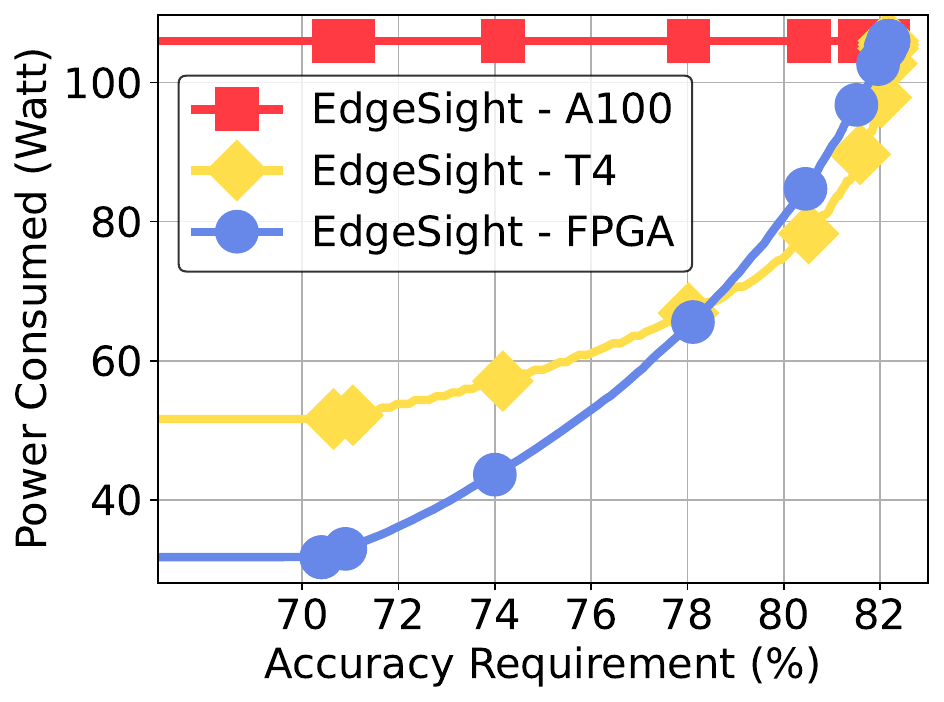}
            	\caption{Power}
                \label{fig:fpga-performance-per-watt}
            \end{subfigure}
            \begin{subfigure}[t]{0.32\textwidth}
                \centering
                \includegraphics[width=\linewidth]{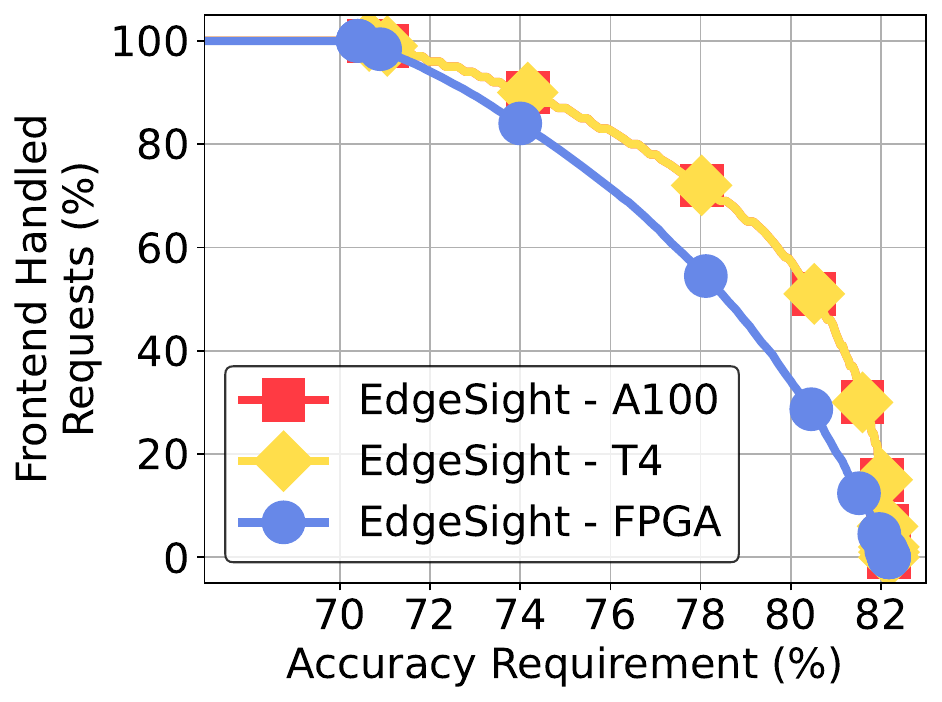}
            	\caption{Percentage of Frontend-only Requests}
                \label{fig:fpga-percentage}
            \end{subfigure}
            \caption{Different hardware with different latency and power consumption.}
    	\end{minipage}
            \vspace{-6mm}
    \end{figure*}
\else
    \begin{figure*}[t]
        \begin{minipage}{0.25\textwidth} 
            \centering
            \includegraphics[width=\linewidth]{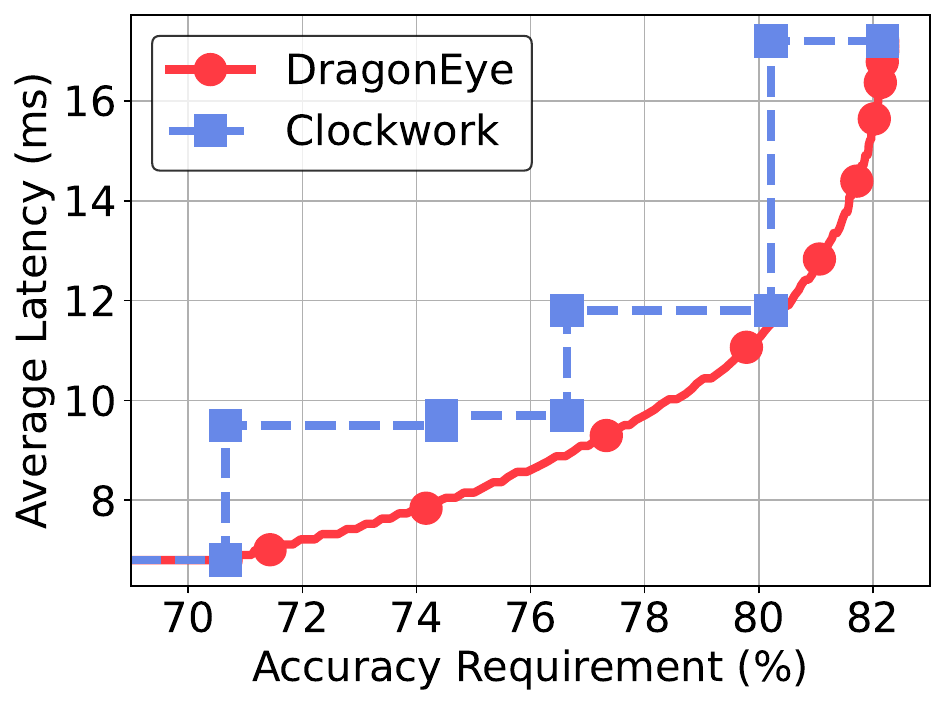}
        	\caption{Latency curve with different accuracy requirement}
            \label{fig:latency_curve}
        \end{minipage}
        \begin{minipage}{0.73\textwidth}
            \begin{subfigure}[t]{0.32\textwidth}
                \centering
                \includegraphics[width=\linewidth]{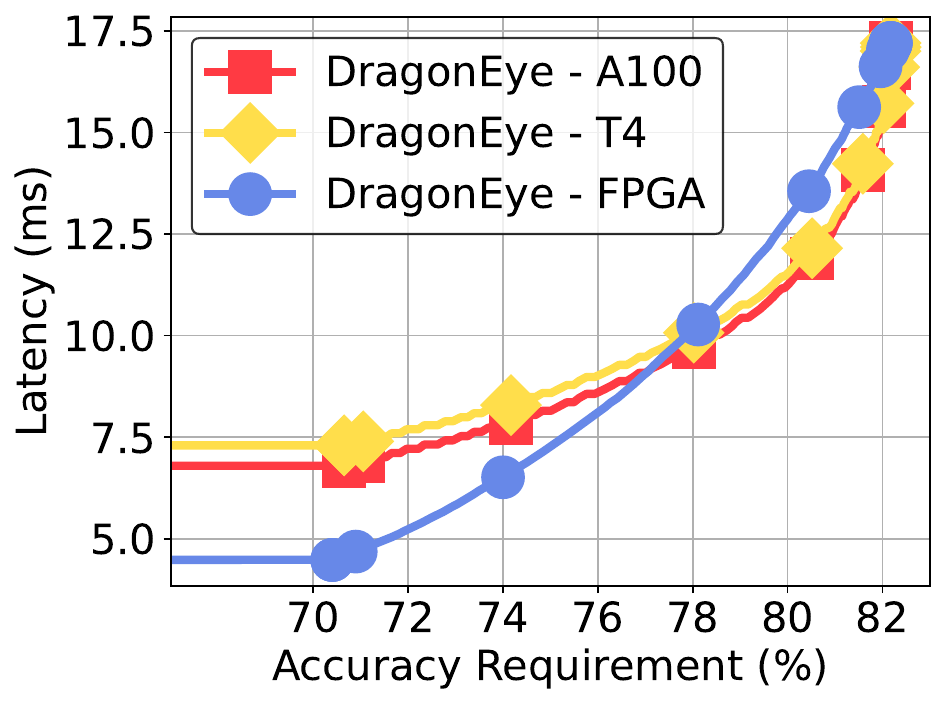}
            	\caption{Latency}
                \label{fig:fpga-performance}
            \end{subfigure}
            \begin{subfigure}[t]{0.32\textwidth}
                \centering
                \includegraphics[width=\linewidth]{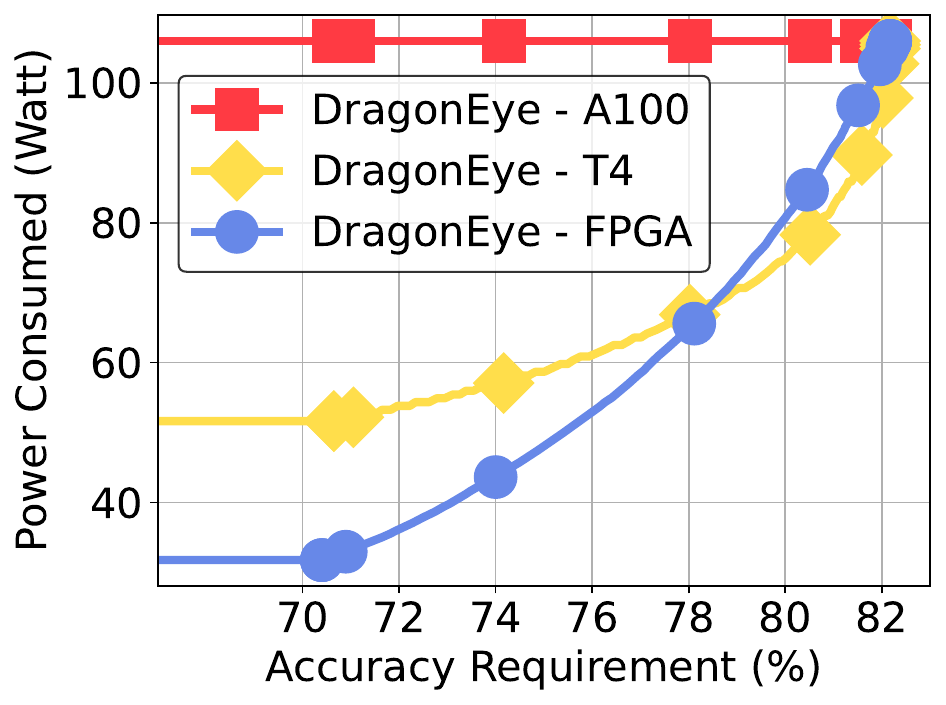}
            	\caption{Power}
                \label{fig:fpga-performance-per-watt}
            \end{subfigure}
            \begin{subfigure}[t]{0.32\textwidth}
                \centering
                \includegraphics[width=\linewidth]{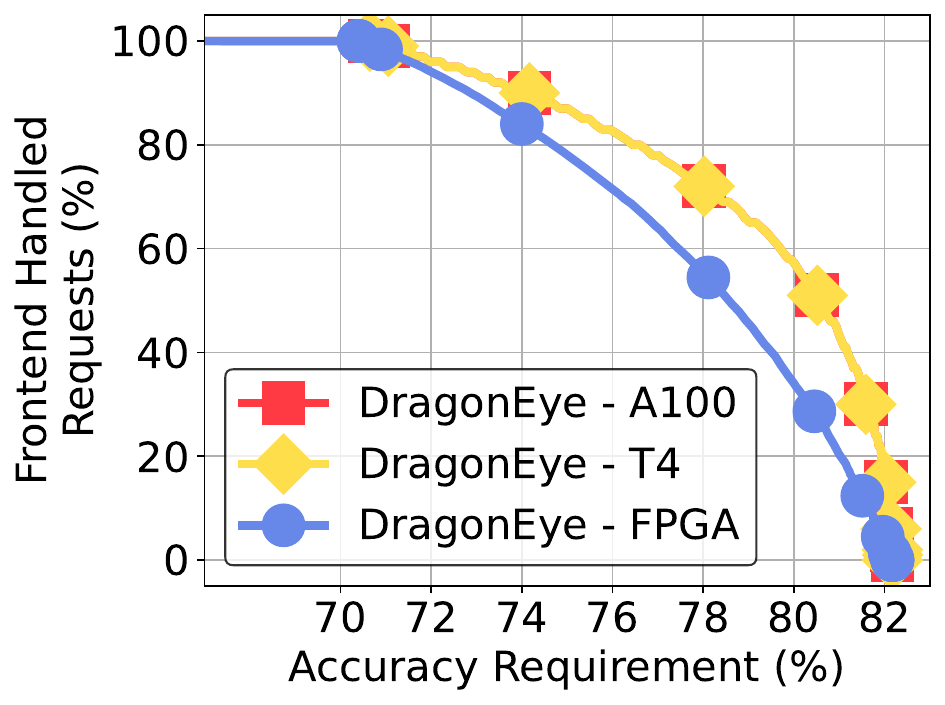}
            	\caption{Percentage of Frontend-only Requests}
                \label{fig:fpga-percentage}
            \end{subfigure}
            \caption{Different hardware with different latency and power consumption.}
    	\end{minipage}
            \vspace{-6mm}
    \end{figure*}
\fi


\subsection{\sysname Uses Request-level Granularity}
\vspace{-1mm}
\label{eval:edge_dc}

We take a deeper look into the  the latency for different accuracy requirements using request-level granularity (\sysname) and model-level granularity (Clockwork). For Clockwork, system selects the model with lowest accuracy that could serve the accuracy requirement from the user request.

For Clockwork, we select 5 models with the accuracy  identical to previous work~\cite{Cocktail}. For \sysname, we select the highest accuracy and lowest accuracy models among the five: $winograd\_resnet18\_v2$ and $resnest101$.

Figure~\ref{fig:latency_curve} demonstrates the latency benefits. In \sysname, latency varies according to different accuracy requirements because different proportions of requests are sent to the backend with confidence scaling. When frontend inference confidence is insufficient, requests fall back to the backend GPU. For instance, at an accuracy requirement of 75\%, approximately 96\% of requests are replied to by the edge vision box in \sysname. However, in a separate model deployment scenario (Clockwork), all requests below a certain accuracy requirement are sent to the closest model which can serve the target accuracy requirement, resulting in high overhead due to serving with overkill models, causing a $0.84X$ worse performance compared to \sysname.

Confidence scaling requires only two models with their confidence scores presented. Since the edge is a relatively power-constrained environment, deploying power-hungry hardware like the A100 is challenging. Therefore, both performance and power are important considerations. To address this issue, we implemented an FPGA prototype. Because \textit{winograd\_resnet18\_v2} cannot be directly deployed to the FPGA, we selected MobileNet, compiled using Xilinx FINN~\cite{finn}, which achieves the same accuracy for a fair comparison.

In Figure~\ref{fig:fpga-performance}, we show that the FPGA achieves similar performance to the A100 and T4 at the edge device, and outperforms the A100 by $1.5X$ with low requirement applications. The performance benefits come from the image processing pipeline and streaming advantages in FPGA architecture. In Figure~\ref{fig:fpga-performance-per-watt}, we can see that the power consumption when using the A100 for the edge is high, with an average of 106W at a 70\% accuracy requirement, while this number decreases to 31.74W with the FPGA prototype, showing a $3.34X$ reduction in power consumption with the FPGA. The percentage of requests handled by the frontend for different accuracy requirements is shown in Figure~\ref{fig:fpga-percentage}.


\subsection{\sysname Handles Packet Loss}
\label{sec:packet_loss}
\begin{figure*}[t]
    \begin{minipage}[t]{0.43\textwidth}
        \centering
        \begin{subfigure}[t]{0.49\linewidth}
            \centering
            \includegraphics[width=\linewidth]{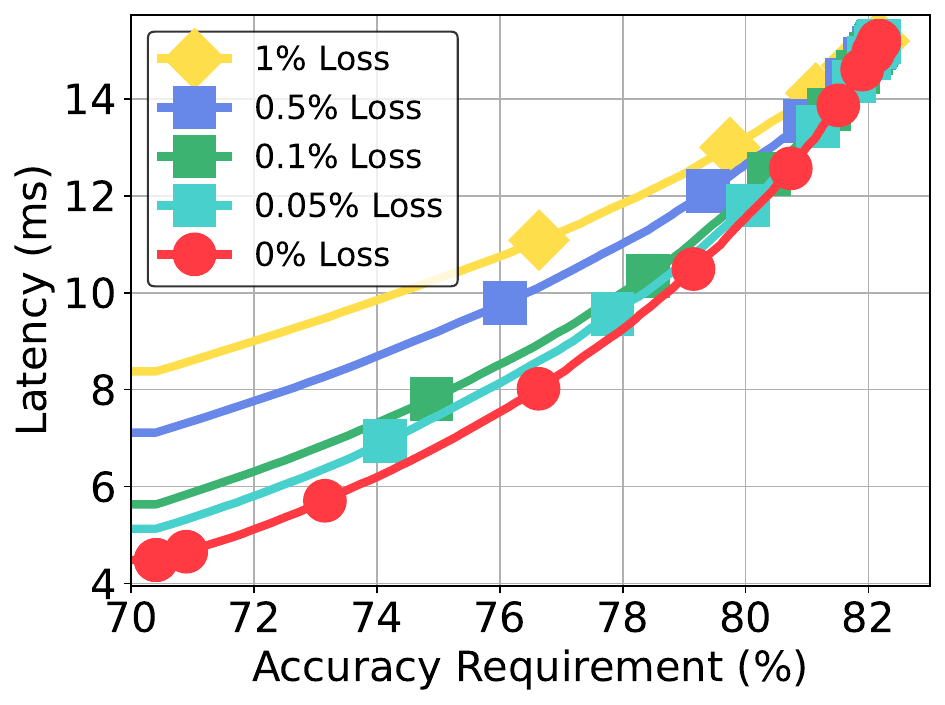}
            \caption{Latency}
            \label{fig:loss_latency}
        \end{subfigure}
        \begin{subfigure}[t]{0.49\linewidth}
            \centering
            \includegraphics[width=\linewidth]{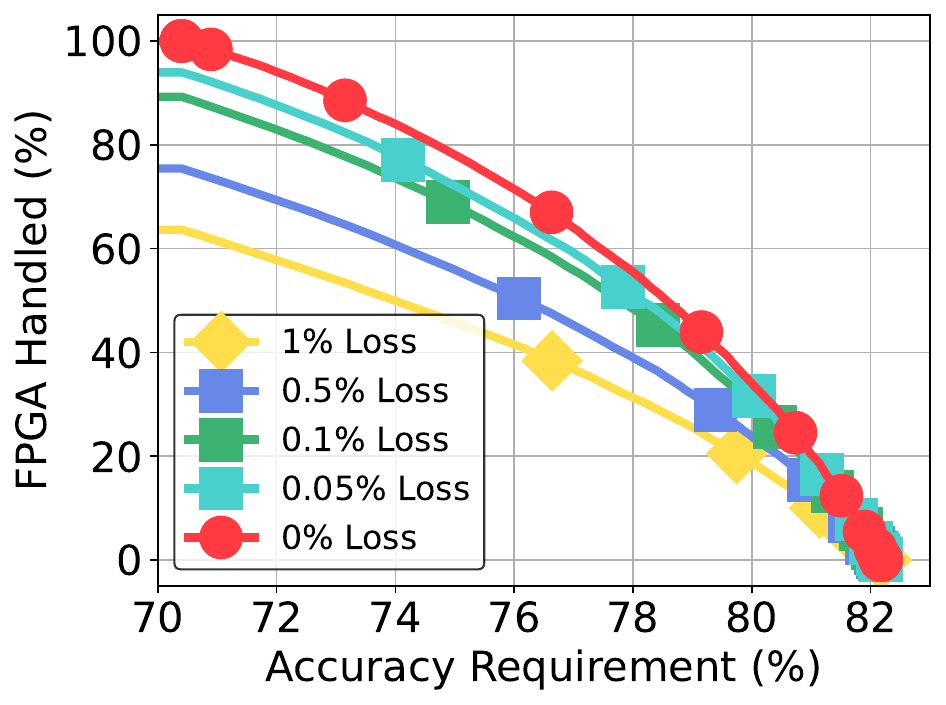}
            \caption{Percentage of request that FPGA Handled}
            \label{fig:fpga_handled}
        \end{subfigure}
        \caption{Encountering different packet loss rate.}
    \end{minipage}%
    \hspace{5pt}
    \begin{minipage}[t]{0.49\textwidth}
        \vspace{-8em}  
        \centering
        \footnotesize 
        \begin{tabular}{|c||c|c||c|c|} 
        \hline
         & \multicolumn{2}{c||}{\textbf{OpenNIC}} & \multicolumn{2}{c|}{\textbf{\sysname}} \\
         \hline
        \textbf{Resources} & \textbf{Percentage} & \textbf{Numbers} &\textbf{Percentage} & \textbf{Numbers} \\
        \hline
        LUT&  6\%& 99638&32\%& 328282  \\
        \hline
        LUTRAM&  2\%& 13781& 8\%& 65563 \\
        \hline
        FF&  4\%& 144632& 12\%& 423430 \\
        \hline
        BRAM&  7\%& 178&60\%& 1620.5 \\
        \hline
        URAM&  1\%& 10&3\%& 32 \\
        \hline
        DSP&  0\%& 0&3\%& 156 \\
        \hline
        \end{tabular}
        \captionof{table}{FPGA resource usage of \sysname and the NIC skeleton}
        \label{tab:resource_usage}
    \end{minipage}
    \vspace{-5mm}
\end{figure*}

Packet loss causes long network delays and significantly impacts performance. \sysname designs a hardware-friendly, streaming-based JPEG recovery scheme that enables lossy inference when packet loss occurs. Figures~\ref{fig:loss_latency} and~\ref{fig:fpga_handled} demonstrate that when \sysname-FPGA encounters various levels of packet loss, it can still function effectively, perform recovery, and make lossy inferences. Our confidence scaling design allows unconfident results due to lossy inference to fall back to the DC backend for handling requests. When 1\% packet loss is encountered, the overall latency increases from 4.48ms to 8.2ms, representing approximately a $1.83X$ degradation. This increase is because, while we can handle lossy inference, achieving the same accuracy for client service without quality loss requires more requests (37\% of the requests for 75\% accuracy) to fall back to the backend for inference.

\subsection{Resource Utilization of \sysname-FPGA}
We now report \sysname power usage compared to GPU solutions and \sysname resource usage on FPGA.

{\bf Power Consumption.}
Our system has lower power consumption compared to GPU solutions. \sysname consumes only 31.74W when it is active. In comparison, an A100 GPU consumes an average of 56W when idle, and this number increases to an average of 106W when making inference requests with the same configuration and a batch size of 1. At maximum load, the power consumption can go up to 250W-400W~\cite{a100}.

{\bf FPGA Resource utilization}
\label{sec:fpga_utilization}
Table~\ref{tab:resource_usage} shows the FPGA resource utilization of \sysname and OpenNIC in our implementation on Xilinx Alveo U250 FPGA Card. The percentages and numbers on the table represent the usage of different types of resources on the FPGA.

The OpenNIC provides the basic network functions to drive the FPGA to send and receive packets with the QSFP28 network interface. The whole \sysname workflow is developed inside the OpenNIC skeleton. Our system uses 26\% more LUT, 8\% more FF, and 53\% more BRAM compared to OpenNIC.\

The table also indicates that we still have available space to potentially add additional functions to the FPGA, introduce other models, or migrate our entire design to a low-end FPGA for better cost savings. This is feasible given that CV models are typically small, and advancements in quantization and sparsification techniques continue to reduce the size of the models. As mentioned in Xilinx FINN~\cite{finn_github}, a VGG-16-like model can be accommodated on a Pynq-Z1 board to further reduce power consumption.




\section{Related Work}

{\noindent \bf Minimizing Inference System Cost.}
Recent works propose model scheduling and resource scaling on GPUs based on SLO requirements~\cite{clockwork, INFaaS}, rather than relying on static provisioning~\cite{olston2017tensorflowserving, triton}. Other approaches include serving inferences on spot instances with fault tolerance mechanisms~\cite{Cocktail}, or sacrificing model accuracy during high demand~\cite{ahmad2024proteus}. These approaches are orthogonal to \sysname, as \sysname reduces the number of model instances in these settings.


{\noindent \bf Hardware Device Acceleration for ML.}
Many hardware solutions have been proposed for accelerating inference applications such as FPGAs~\cite{blott2018finn, finn, vitis_ai, DBLP:conf/icfpt/PetricaAKFCB20,DBLP:journals/corr/abs-2007-10451,10.1145/3404397.3404473,10.1145/3289602.3293915,8825027,10.1145/3330345.3330385,10.1145/3330345.3330385,10.1145/3289185,yolo2,MLSYS2021_PANAMA}. N3IC~\cite{n3ic} uses a SmartNIC to help machine learning inference. Taurus~\cite{taurus}, SwitchML~\cite{switchml265065}, ATP~\cite{lao2021atp} and other works~\cite{sanvito2018can, hotnets19dream} discuss the possibility offload machine learning applications to the network programmable switches. None of these approaches consider a middlebox-style approach where FPGA adds functionality to an end-to-end inference pipeline.

{\bf Reducing both network and data processing taxes.} 
While previous serving system work has primarily focused on improving model inference times through better resource scheduling within data centers~\cite{clipper, INFaaS, clockwork, SHEPHERD, Cocktail}, or optimizing the models themselves~\cite{DBLP:conf/icfpt/PetricaAKFCB20,DBLP:journals/corr/abs-2007-10451,10.1145/3404397.3404473,10.1145/3289602.3293915,8825027,10.1145/3330345.3330385,10.1145/3330345.3330385,10.1145/3289185,finn, blott2018finn}, a significant portion of the end-to-end inference delay stems from network and data processing overhead, which our work addresses but previous systems have not considered.


\vspace{-1mm}
\section{Discussion and Future Work}

While this paper focuses on vision models, the proposed techniques (e.g., confidence scaling, lossy inference) are generalizable and applicable to other domains like language models or multimodal systems.

\mypar{Applies \sysname to Language-based Tasks. }
\label{dis:language}
While \sysname is demonstrated with vision models in this paper, the contributions apply equally well to language processing models. Recent research~\cite{xiong2024can,li2024think} has explored the confidence of answers output from large language models (LLMs) by instructing the models to output the confidence or the quality of the results. The findings are similar to those in \secref{sec:confidence_threshold}, where higher confidence corresponds to better evaluation results. Moreover, although there are no publicly available LLM inference deployments for FPGA at this time, recent developments in deploying LLMs to FPGA for inference are making progress~\cite{fpgallm}.
With these advancements, the FPGA frondend and the backend server can adapt to different tasks, following the hierarchical design of \sysname. By adjusting the confidence threshold, \sysname can control the quality or accuracy of tasks, ensuring flexibility and efficiency across various applications.

\mypar{Applies \sysname to Multimodal. }The growing use of multimodal services, such as Visual Question Answering (VQA) and image captioning, offers a valuable opportunity for \sysname to adapt and integrate. These systems often face imbalances~\cite{distmm}, as image and text data are processed separately, with vision tasks frequently becoming a bottleneck in vision-intensive applications. Applying \sysname to the vision data can reduce inference time, improve integration with text processing, and accelerate overall performance in multimodal systems.

\mypar{Inference Model on FPGA. }
\label{dis:fpga}
In recent years, there have been many works~\cite{DBLP:conf/icfpt/PetricaAKFCB20,DBLP:journals/corr/abs-2007-10451,10.1145/3404397.3404473,10.1145/3289602.3293915,8825027,10.1145/3330345.3330385,10.1145/3330345.3330385,10.1145/3289185} proposed to accelerate inference using FPGAs. There has been a range of advances that have progressively packed complex models on to FPGAs with low memory footprint and low latency,  while minimally compromising accuracy.
For example, prior research~\cite{Binarized_Neural_Networks,kim2016bitwise,rastegari2016xnor,NIPS2016_bnn,Cai_2017_CVPR,bnn_survey,Chen_2021_CVPR,10.1007/978-3-030-58536-5_5} 
shows that applying quantization to models can greatly accelerate FPGA inference speed via simplified models, 
while still maintaining a good level of accuracy, even when using a binary representation of the parameters. 
Furthermore, previous work has shown that we can remove the redundant parameters that do not contribute much to 
the accuracy~\cite{NIPS2015_ae0eb3ee,10.1145/3297858.3304028,DBLP:journals/corr/LiKDSG16,DBLP:journals/corr/abs-1803-03635}.
This can not only improve inference throughput, but also reduce the model's memory footprint 
and allow the FPGA to host a more sophisticated model(s).
Solutions such as~\cite{10.1145/3330345.3330385, 10.1145/3297858.3304028} further accelerate the 
FPGA inference pipeline by converting the model into a streamlined architecture. 
Furthermore, operator fusion~\cite{blott2018finn, finn} can be used to merge  complex operators (\emph{e.g.} floating-point operations) at compile time for even greater speed. 
Our work can leverage all these advances when hosting models on the FPGA with the goal of keeping the FPGA model's accuracy close to the GPU-resident model, yet simple so as to yield tangible memory and performance improvements. No matter the advances, there will likely be an FPGA-GPU model accuracy gap, so we need an approach to bridge the outcome of that gap.



\vspace{-1mm}
\section{Conclusion}

We built \sysname, an edge-DC system where the frontend supports compressed format recovery and lossy inference in client requests. It also applies confidence scaling with the backend to support a wide range of accuracy requirements, achieving better performance gains and reducing deployment costs. Our FPGA prototype also demonstrates the feasibility of implementing this system in a more power-efficient way.

\bibliographystyle{plain}
\bibliography{reference}
\appendix
\end{document}